\documentclass[a4paper,fleqn,usenatbib]{mnras}
\usepackage{lscape,graphicx}
\usepackage{rotating}
\usepackage{color}
\usepackage{threeparttable}
\usepackage{amsmath,amssymb}

\usepackage{color}

\def\apj{ApJ}
\def\apjs{ApJS}
\def\mnras{MNRAS}
\def\nat{Nat}
\def\apjl{ApJ}
\def\aap{A\&A}
\def\aj{AJ}

\def\araa{ARA\&A}

\def\asec{\ifmmode ^{\prime\prime}\else$^{\prime\prime}$\fi}

\def\it{\sl}
\def\degs{\ifmmode ^{\circ}\else$^{\circ}$\fi}
\def\amin{\ifmmode ^{\prime}\else$^{\prime}$\fi}
\def\asec{\ifmmode ^{\prime\prime}\else$^{\prime\prime}$\fi}
\def\fdg{\hbox{$.\!\!^\circ$}}          

\def\degs{\ifmmode ^{\circ}\else$^{\circ}$\fi}
\def\amin{\ifmmode ^{\prime}\else$^{\prime}$\fi}

\def\eqalign#1{\null\,\vcenter{\openup1\jot \m@th
   \ialign{\strut\hfil$\displaystyle{##}$&$\displaystyle{{}##}$\hfil
   \crcr#1\crcr}}\,}
\def\G350{G350.0$-$2.0}

\title[\textit{XMM-Newton} Studies of the SNR G350.0$-$2.0 ]{\textit{XMM-Newton} Studies of the Supernova Remnant G350.0$-$2.0 }
\author[A.~Karpova, P.~Shternin, D.~Zyuzin, A.~Danilenko, Yu.~Shibanov]{
A.~Karpova$^{1,2}$\thanks{E-mail: annakarpova1989@gmail.com},
P.~Shternin$^{1}$, 
D.~Zyuzin$^{1}$, 
A.~Danilenko$^{1}$, 
Yu.~Shibanov$^{1,2}$\\
$^{1}$Ioffe Institute, Politekhnicheskaya 26, St. Petersburg, 194021, Russia\\
$^{2}$Peter the Great St. Petersburg Polytechnic University, Politekhnicheskaya 29, St. Petersburg, 195251, Russia}

\date{Accepted XXX. Received YYY; in original form ZZZ}
\pubyear{2016}

\begin{document}

\label{firstpage}
\pagerange{\pageref{firstpage}--\pageref{lastpage}} 
\maketitle

\begin{abstract}   
We report the results of \textit{XMM-Newton} observations of the Galactic
mixed-morphology supernova remnant \G350.
Diffuse thermal X-ray emission fills the north-western part 
of the remnant surrounded by radio shell-like structures.
We did not detect any X-ray counterpart of the latter structures,
but found several bright blobs within the diffuse emission.
The X-ray spectrum of the most part of the remnant 
can be described  by a collisionally-ionized
plasma model {\sc vapec} with solar abundances and 
a temperature of $\approx 0.8$ keV. 
The solar abundances of plasma indicate that the X-ray emission comes
from the shocked interstellar material.
The overabundance of Fe was found in some of the bright blobs.
We also analysed the brightest point-like X-ray source 1RXS J172653.4$-$382157
projected on the extended emission.
Its spectrum is well described by the two-temperature
optically thin thermal plasma model {\sc mekal} typical for cataclysmic variable stars. 
The cataclysmic variable source nature is supported by the presence of a faint ($g\approx21$) 
optical source with non-stellar spectral energy distribution
at the X-ray position of 1RXS J172653.4$-$382157.
It was detected with the \textit{XMM-Newton} optical/UV monitor in the $U$ filter 
and was also found in the archival H$\alpha$ and 
optical/near-infrared broadband sky survey images.
On the other hand, the X-ray spectrum is also described  
by the power law plus thermal component model
typical for a rotation powered pulsar.
Therefore, the pulsar interpretation of the source cannot be excluded.
For this source, we derived the upper limit for the pulsed fraction of 27 per cent.

\end{abstract}


\section{Introduction}
Mixed-morphology (MM) supernova remnants (SNRs) are 
characterized by a shell-like morphology in the radio 
and a centrally filled thermal emission in X-rays \citep[e.g.,][]{rho1998}.
This class represents about 8 per cent of total Galactic SNR population 
and about 25 per cent of all Galactic SNRs observed in X-rays \citep{rho1998}.
Two main scenarios were proposed to explain MM SNRs morphology:
the first one is based on effects of thermal conduction processes
within the SNR interior \citep[e.g.,][]{cox1999} 
and the other one -- on a cloudlet evaporation \citep[e.g.,][]{white1991}
assuming the SNR shock propagation through a cloudy ISM.
A fraction of MM SNRs shows enhanced 
metal abundances \citep[see, e.g.,][]{lazendic2006}
that is not properly addressed by traditional models.
Determination of MM SNRs properties, such as
temperatures, abundances and densities,
is important to improve the models of their formation.  

\begin{figure*}
\begin{minipage}[h]{0.49\linewidth}
\center{\includegraphics[width=0.87\linewidth,clip]{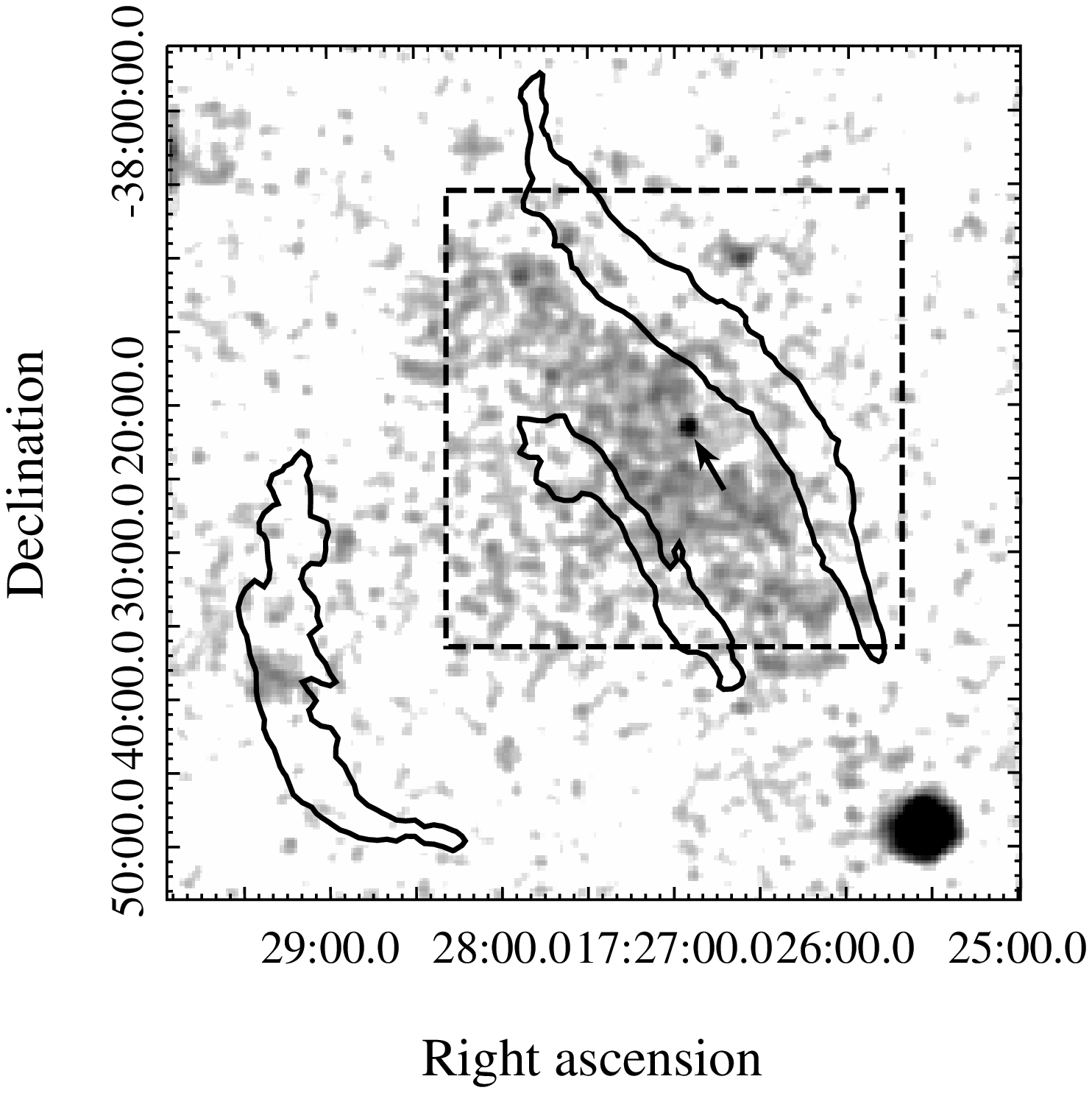}}
\end{minipage}
\begin{minipage}[h]{0.49\linewidth}
\center{\includegraphics[width=0.94\linewidth,clip]{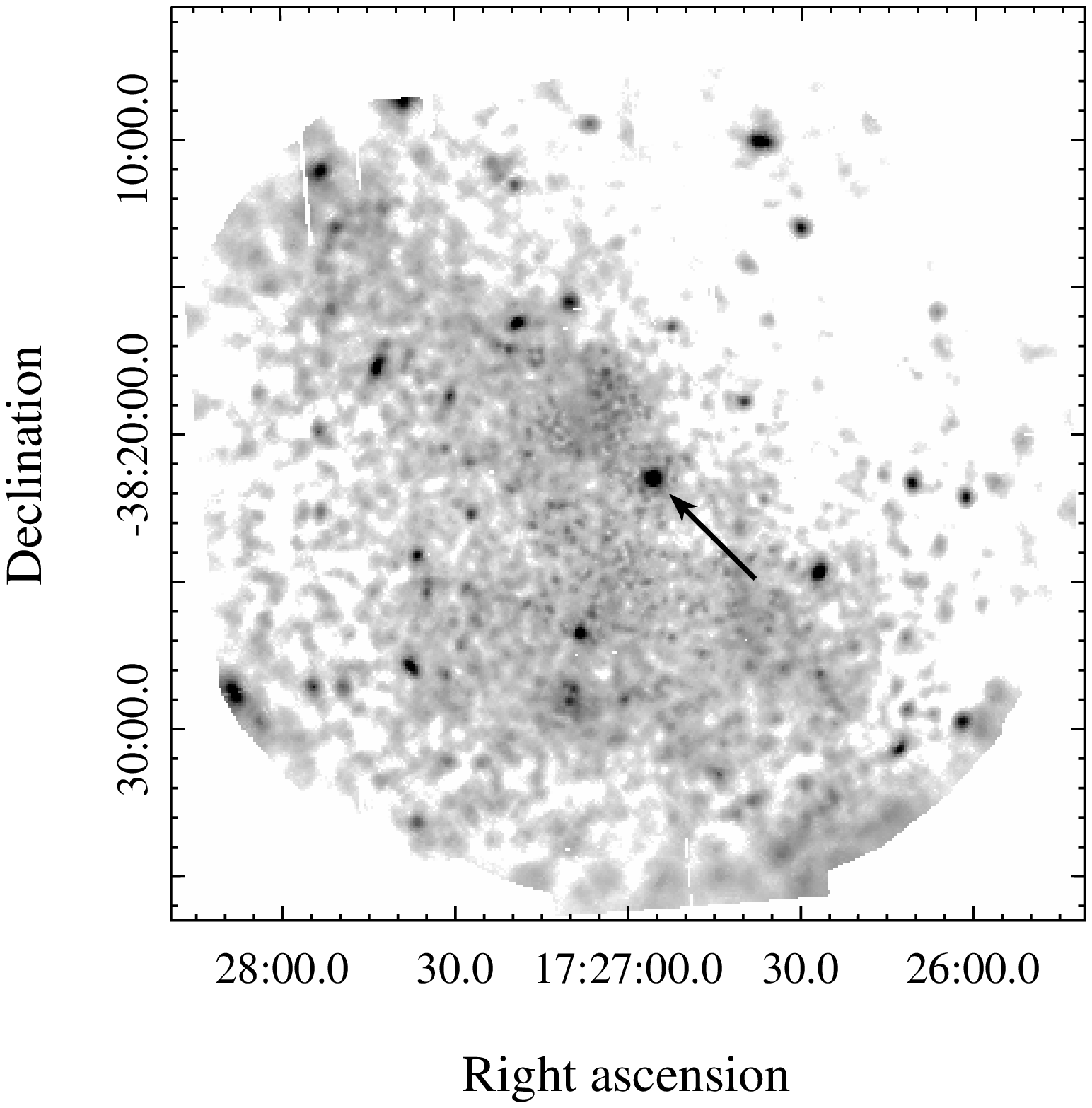}}
\end{minipage}
\caption{\textit{ROSAT} 58 arcmin$\times$58 arcmin 
\G350 field in 0.1--2.4 keV band (left panel) 
together with the VLA 1.4 GHz radio contours \citep{gaensler1998}.
31 arcmin$\times$31 arcmin exposure and vignetting corrected
QPB-subtracted \textit{XMM-Newton} image in 0.4--7.2 keV band is
shown in the right panel (the data of MOS and pn detectors are combined).
The respective region is shown in the left panel with the dashed square.
Square root brightness scale is used for both images.
The X-ray source J1726 is marked by the arrows.}
\label{fig:img}
\end{figure*}

\begin{figure*}
\begin{minipage}[h]{0.49\linewidth}
\center{\includegraphics[width=0.83\linewidth,clip]{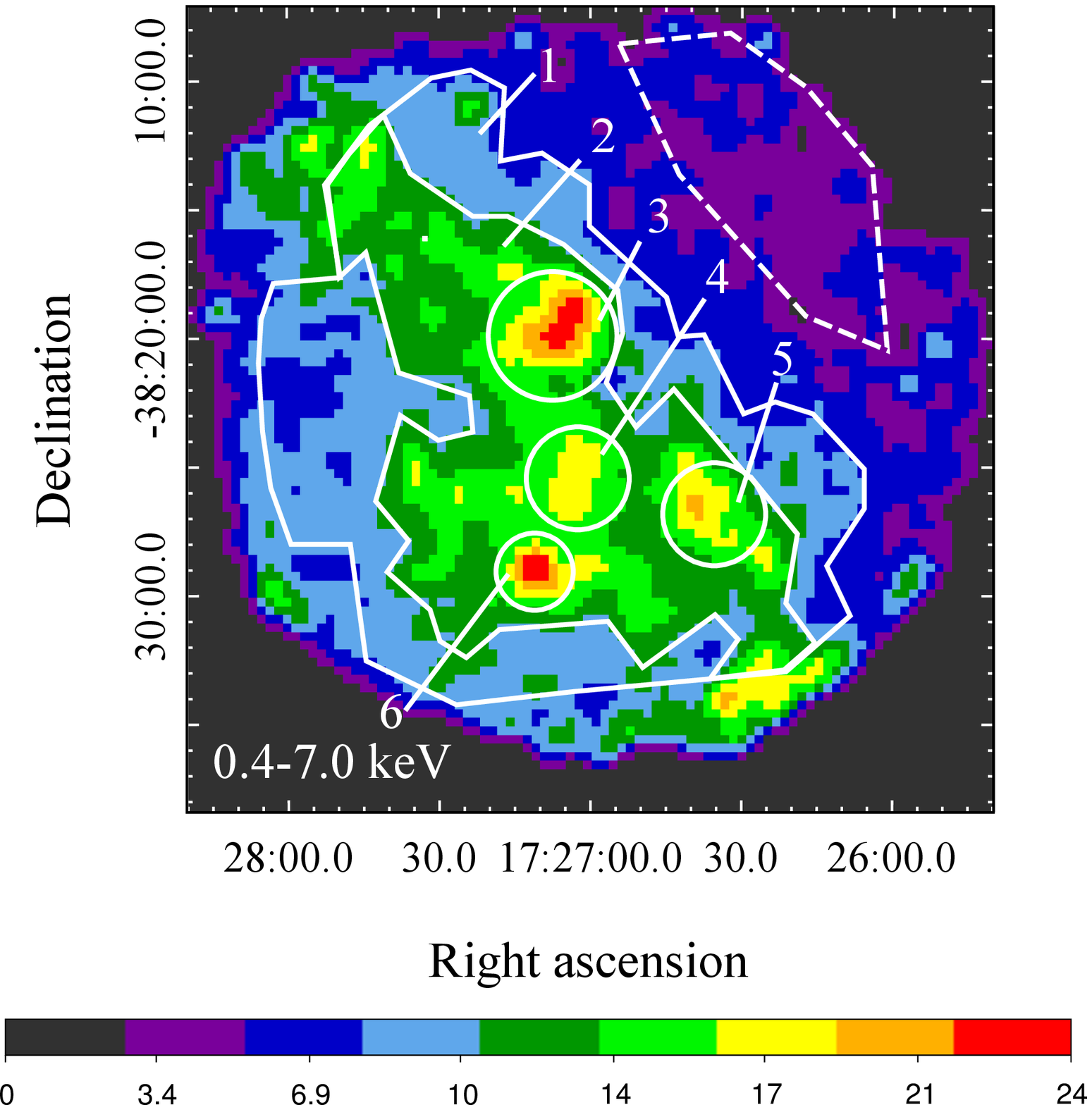}}
\end{minipage}
\begin{minipage}[h]{0.49\linewidth}
\center{\includegraphics[width=0.83\linewidth,clip]{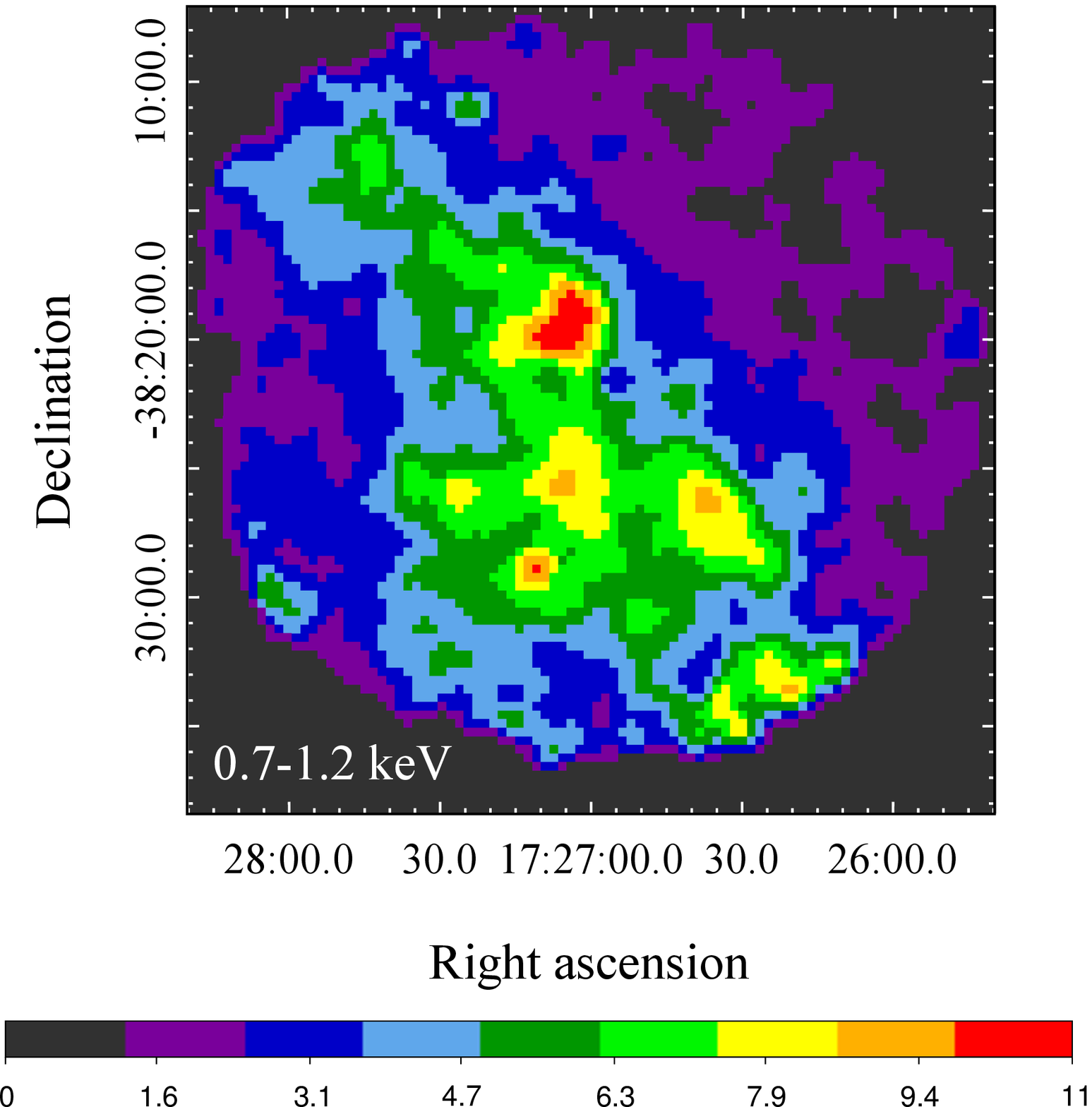}}
\end{minipage}
\begin{minipage}[h]{0.49\linewidth}
\center{\includegraphics[width=0.83\linewidth,clip]{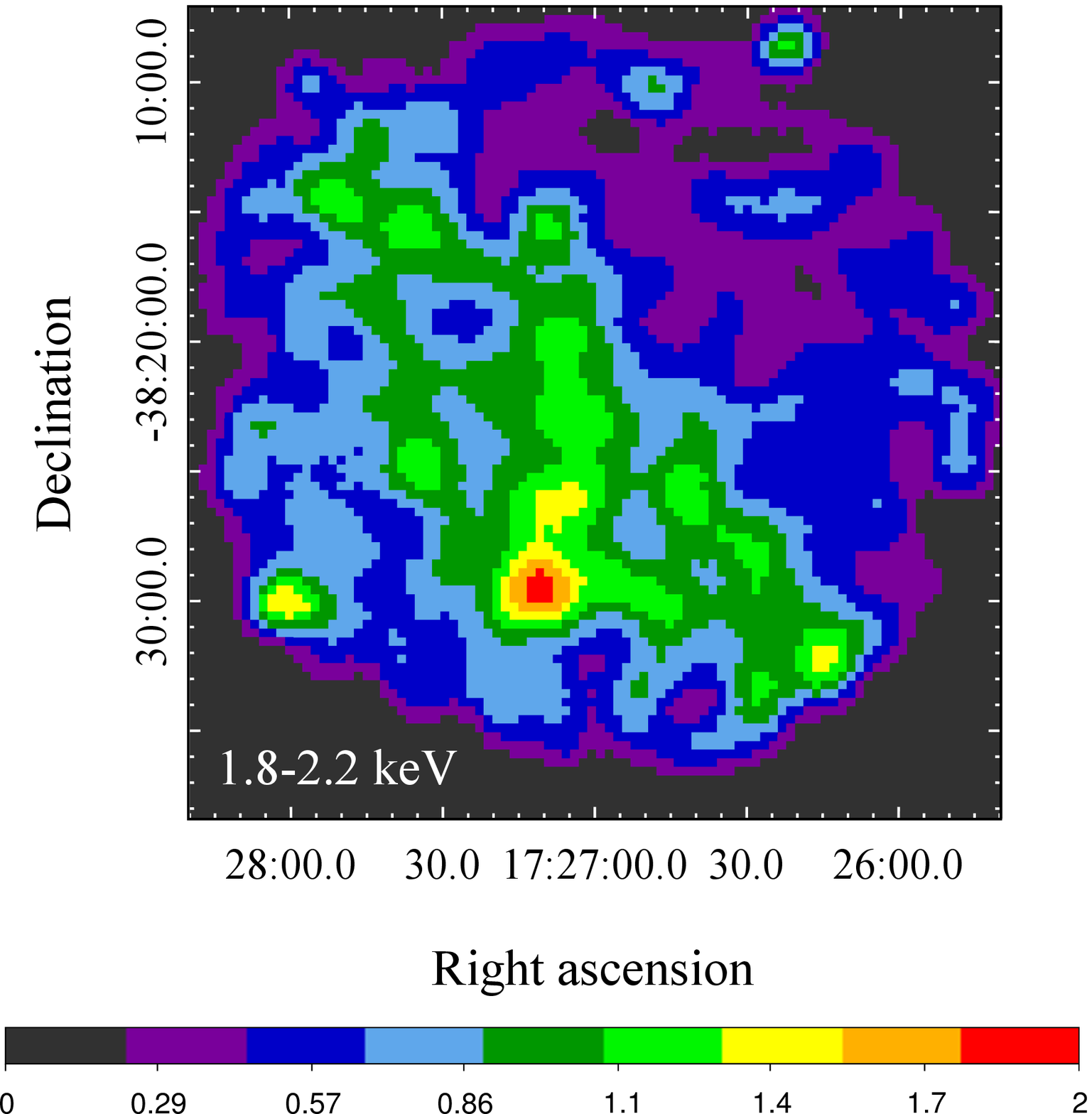}}
\end{minipage}
\begin{minipage}[h]{0.49\linewidth}
\center{\includegraphics[width=0.83\linewidth,clip]{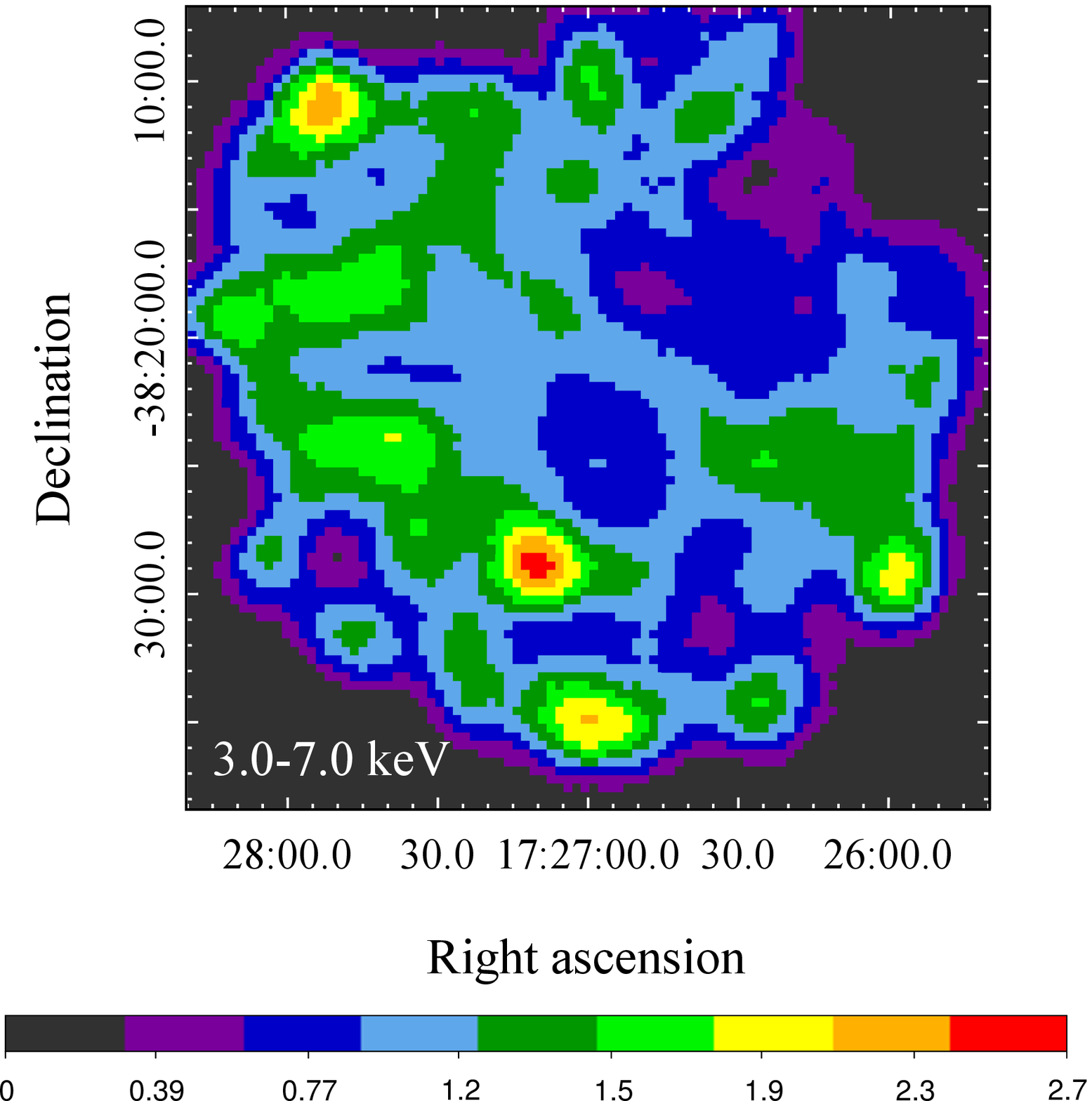}}
\end{minipage}
\caption{\G350~images in 0.4--7 keV (top left), 0.7--1.2 keV (top right), 1.8--2.2 keV (bottom left) and 3--7 keV (bottom right) energy bands. The data of MOS and pn detectors are combined. 
The images are exposure and vignetting corrected and QPB-subtracted. 
Point sources are removed. The pixel size is 20 arcsec.
The intensity is given in counts s$^{-1}$ deg$^{-2}$. 
In the top left panel, regions used for spectral analysis are shown (see text for details).}
\label{fig:snr_images}
\end{figure*}


SNR \G350~was discovered in the radio 
with Molonglo and Parkes telescopes at 408 MHz and 5 GHz, respectively \citep{caswell1975}.
Very Large Array (VLA) observations \citep{gaensler1998} showed 
complicated morphology of \G350 that consists of 
three spatially distinct emission regions:
the bright north-western and fainter inner and south-eastern arcs 
(Fig.~\ref{fig:img}, left panel).
The whole extent of \G350 in the radio is $\sim 40$ arcmin.
In the optical, \citet{stupar2011} found some H$\alpha$ filaments and clumps
spatially coinciding with the radio structures.
\G350~was observed in X-rays with \textit{ROSAT} and \textit{ASCA}.
The multiwavelength data show that the SNR belongs to the MM SNR type 
(see Fig.~\ref{fig:img}, left panel). 
The age of the SNR was estimated as
$\sim 10^4$ yr assuming the Sedov expansion phase \citep*{helfand1980,clark1975}.
Basing on the radio surface brightness-to-diameter relationship, \citet{case1998} estimated the distance $D$ to \G350~to be about 3.7 kpc. 

A bright X-ray point-like source, 1RXS J172653.4$-$382157 (hereafter J1726), 
was detected in the SNR field with \textit{ROSAT} 
(see Fig.~\ref{fig:img}, left panel) and \textit{ASCA}.
It may be an associated neutron star (NS) 
although no radio pulsar was detected within the SNR \citep{kaspi1996}.
On the other hand, it may be an unrelated object.  
To study the SNR and J1726 properties, we performed  
\textit{XMM-Newton}\footnote{PI Zyuzin, \textit{XMM-Newton}/EPIC, ObsID 0724220101} observations. 
Here we present results of the analysis of the \G350 and J1726 X-ray emission. 
The details of observations and imaging are described in Section~\ref{sec:data}.
The analysis of the X-ray SNR spectra is presented in Section~\ref{sec:spec}.
J1726 is analysed in Section~\ref{sec:j1726}.
We discuss results in Section~\ref{sec:discus}
and summarise them in Section~\ref{sec:sum}.

\section{X-ray data and imaging}
\label{sec:data}

The \textit{XMM-Newton} observations of \G350 were 
carried out on 2013 September 21 with total exposure of about 38 ks. 
The telescope was pointed to the J1726 position.
The EPIC field of view covers almost
whole diffuse emission seen with \textit{ROSAT} in the north-western part of the SNR
(Fig.~\ref{fig:img}, left panel).
All EPIC cameras were operated in the Full Frame Mode 
with the medium filter setting. 
The {\sc xmm-sas}~v.13.5.0 software was used to analyse the data.
Data were reprocessed using the SAS tasks
{\sc emchain} and {\sc epchain}.

We used the \textit{XMM-Newton} Extended Source Analysis Software\citep[{\sc esas};][]{cookbook}
to analyse the SNR emission. 
Utilizing {\sc mos-filter} and {\sc pn-filter} tools, 
we filtered out bad time intervals caused by soft proton flares. 
The resulting exposures are 19.5, 21.3 and 15.8 ks 
for EPIC-MOS1, MOS2 and pn, respectively. 
To identify whether any MOS CCD was in an
anomalous state, we ran {\sc emtaglenoise} task.
MOS2 CCD \#5 showed enhanced background below 1 keV 
and was excluded from further analysis. 
MOS1 CCDs \#3 and \#6 were damaged by 
micrometeorite strikes and are no longer 
functional \citep{cookbook}.
Therefore, MOS1 does not cover the entire SNR region
and was not used for the spectral analysis of \G350 emission.
We selected single and double pixel events 
(PATTERN~$\leq$~4) for the EPIC-pn 
and single to quadruple-pixel events 
(PATTERN~$\leq$~12) for the EPIC-MOS data. 
Images were created by {\sc mos-spectra} and {\sc pn-spectra} tasks. 
{\sc mos\_back} and {\sc pn\_back} routines were run 
to create the quiescent particle background (QPB) images.
Then MOS and pn QPB-subtracted images were combined and adaptively smoothed.
The resulting image in 0.4--7.2 keV energy range is shown in 
the right panel of Fig.~\ref{fig:img}.
The shape of diffuse emission is generally consistent with that seen
in the \textit{ROSAT} image presented in the left panel.
However, deeper \textit{XMM-Newton} observation
revealed many point-like sources in the SNR field.
We also constructed images for 0.4--7, 0.7--1.2, 1.8--2.2 and 3--7 keV bands
to investigate spectral variations of the SNR morphology.  
In these images, point-like sources were removed using {\sc cheese} tool
and MOS and pn images were combined and binned using {\sc bin\_image} task.
The resulting images are shown in Fig.~\ref{fig:snr_images}.
We did not find any evidence of the X-ray counterparts of the radio shells.
The most prominent structures are blobs 3--6 found in the inner part
of the diffuse emission; they are marked in the top left panel of Fig.~\ref{fig:snr_images}.
Blobs 3--5 reveal themselves brighter in the 0.7--1.2 keV band where the Fe-L  
emission dominates in the SNR spectrum.
Blob 6 seems to have the spectrum different from the rest of the SNR  
since its emission is harder.

\section{Spectral analysis of \G350} 
\label{sec:spec}

\begin{table*}
\caption{Best-fit parameters for the absorbed {\sc vapec} model 
for the SNR emission from different regions. 
Parameters for additional Gaussian component for regions 1--5 and for power law
emission for the region 6 are also given (see text for details).
Errors are at 90\% confidence. }
\label{t:best-fit}
\begin{center}
\begin{tabular}{cccccc}
\hline
Region & 1 & 2 & 3 & 4+5 & 6 \\
\hline
\multicolumn{6}{c}{}\\
Column density $N_{\rm H}$, $10^{21}$ cm$^{-2}$ & $7.1^{+0.5}_{-0.6}$ & $6.2^{+0.3}_{-0.3}$ & $7.2^{+2.0}_{-1.5}$ & $6.6^{+1.4}_{-1.4}$ & $5.3^{+1.6}_{-1.2}$ \\
\hline
\multicolumn{6}{c}{}\\
Temperature $T$, keV & $0.85^{+0.05}_{-0.04}$ & $0.82^{+0.02}_{-0.02}$ & $0.85^{+0.04}_{-0.05}$ & $0.84^{+0.03}_{-0.05}$ & $0.85^{+0.05}_{-0.08}$ \\
\multicolumn{6}{c}{}\\
VEM, $10^{56}D^2_{\rm 3.7kpc}$ cm$^{-3}$\ $^a$   & $3.1^{+0.4}_{-0.3}$ & $4.1^{+0.4}_{-0.4}$ & $0.5^{+0.1}_{-0.1}$ & $0.7^{+0.1}_{-0.1}$ & $0.2^{+0.1}_{-0.1}$ \\
\multicolumn{6}{c}{}\\
Fe/Fe$_\odot$ & 1.0 (fixed) & 1.0  (fixed) & $2.3^{+2.3}_{-1.0}$ & $1.8^{+1.0}_{-0.7}$ & 1.0  (fixed) \\
\hline
\multicolumn{6}{c}{}\\
Gaussian energy $E$, keV & $1.26^{+0.02}_{-0.02}$ & $1.26^{+0.01}_{-0.01}$ & $1.23^{+0.04}_{-0.04}$ & $1.24^{+0.04}_{-0.03}$ & \\
\multicolumn{6}{c}{}\\
Gaussian width $\sigma$, keV & $0.05^{+0.02}_{-0.03}$ & $0.07^{+0.01}_{-0.01}$ & $0.08^{+0.05}_{-0.07}$ & $0.08^{+0.04}_{-0.05}$ & \\
\multicolumn{6}{c}{}\\
Gaussian normalization $K$, $10^{-4}$ ph cm$^{-2}$ s$^{-1}$  & $0.9^{+0.4}_{-0.3}$ & $1.8^{+0.3}_{-0.3}$ & $0.6^{+0.3}_{-0.3}$ & $0.5^{+0.2}_{-0.2}$ & \\
\hline
\multicolumn{6}{c}{}\\
Photon index $\Gamma$ &  &  &  &  & $1.8^{+0.7}_{-1.2}$\\
\multicolumn{6}{c}{}\\
PL normalization $N_{\rm PL}$, $10^{-5}$ ph cm$^{-2}$ s$^{-1}$ keV$^{-1}$ &  & & & & $2.2^{+2.9}_{-1.8}$\\
\hline
\multicolumn{6}{c}{}\\
$\chi^2$/d.o.f.$^b$ & \multicolumn{2}{c}{1063/1001} & 265/261 & 338/320 & 74/67 \\
\hline
\end{tabular}
\end{center}
\begin{tablenotes}
\item $^a$ VEM is the volume emission measure defined in equation~(\ref{eq:norm}) and
$D_{\rm 3.7kpc}$ is the distance in units of 3.7 kpc. 
\item $^b$ d.o.f. = degrees of freedom.
\end{tablenotes}
\end{table*}

We extracted the SNR spectra from MOS2 and pn data
and created redistribution matrix and ancillary response
files using {\sc mos-spectra} and {\sc pn-spectra} tools.
The extraction regions are shown in the top left panel of Fig.~\ref{fig:snr_images}.
The outer region 1 was chosen on the overlapping part of MOS2 and pn fields of view 
between contour levels of 7.5 and 11 cnt s$^{-1}$ deg$^{-2}$. 
The region 2 corresponds to the central part of the remnant with brightness 
$>$11 cnt s$^{-1}$ deg$^{-2}$.
Bright regions 3--6 were excluded from the region 2 and analysed separately. 
MOS2 and pn spectra were grouped to ensure $>50$ and $>20$ counts
per energy bin for regions 1--2 and 3--6, respectively. 
Unfortunately, we could not obtain the astrophysical background spectrum 
for MOS2 data since the only appropriate region
was exposed on CCD \#5 which was in anomalous mode.
Thus, it was extracted only
from the pn data (from the dashed polygon region
depicted in Fig.~\ref{fig:snr_images}, top left panel).
The background spectrum was binned with a minimum of 50 counts per bin.

The spectra were fitted in the 0.4--7 keV range to exclude strong 
pn instrumental Ni-K$\alpha$, Cu-K$\alpha$ and Zn-K$\alpha$ lines
which contaminate harder part of the spectra.
Two Gaussians (model {\sc gauss} in {\sc xspec}) 
with energies of $\sim 1.5$ and $\sim 1.75$ keV were used
to describe Al-K$\alpha$ and Si-K$\alpha$ instrumental lines.
For the astrophysical background, we used the model consisting 
of a cool unabsorbed thermal component with temperature of 0.1 keV
for the emission from the Local Bubble, 
two absorbed thermal components
for the emission from the cooler and hotter Galactic halo, 
and an absorbed power law with $\Gamma=1.46$ describing 
the unresolved background of cosmological sources \citep[see][]{cookbook}.
Additional power law components not convolved with
the instrumental response were utilized to
account for the residual soft proton (SP) contamination.
We set limits on the photon indices of these components at 0.1 and 1.4
following the {\sc esas} Cookbook \citep{cookbook}.
For the interstellar absorption, we used the {\sc xspec} 
photoelectric absorption {\sc phabs} model with
solar abundances of \citet{angr1989}.

\begin{figure}
\begin{minipage}[h]{1.0\linewidth}
\center{\includegraphics[width=0.67\linewidth, angle=-90]{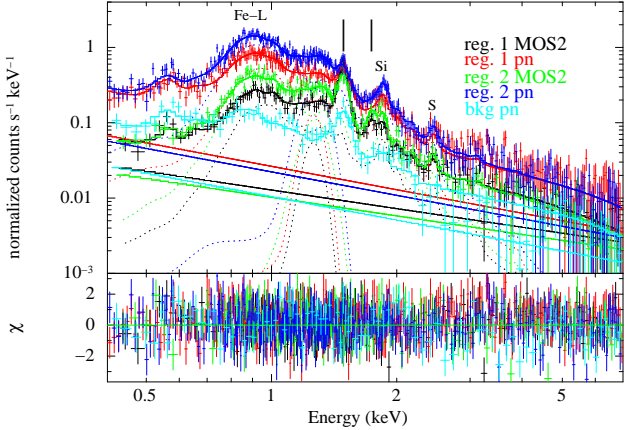}}
\end{minipage}
\caption{Top: \textit{XMM-Newton} EPIC spectra of \G350\ emission from regions 1--2 and the background region. Lines represent the best-fit model, 
assuming {\sc vapec}+{\sc gauss} components for the SNR emission  
(see text for details). 
Solid inclined lines represent the SP contamination, 
dotted lines -- {\sc vapec} and {\sc gauss} components.
Different instruments/regions are shown by different colors
as indicated in the top right corner. 
The instrumental lines positions are shown by thick bars.
Emission features of Fe-L, Si and S are indicated.
Bottom: fit residuals.}
\label{fig:spec}
\end{figure}

For the SNR emission, we used the collisionally-ionized 
equilibrium plasma model {\sc vapec}. 
Spectra of regions 1, 2 and astrophysical background  were fitted
simultaneously. 
The excess in residuals was seen near the energy of 1.25 keV
which we could not remove varying abundances so
we added a Gaussian component to the {\sc vapec} model.
The best-fit parameters are presented in Table~\ref{t:best-fit}
and the best-fit model is shown in Fig.~\ref{fig:spec}.
Parameters of astrophysical background obtained from this fit
were then used for fitting spectra of regions 3--6.

For the bright regions 3--5, 
the {\sc vapec} model with solar abundances does not fit the spectra well.
The best fits (see Table~\ref{t:best-fit}) indicate an overabundance of iron.
As for the regions 1--2, the additional Gaussian component is needed to remove the excess in residuals at 1.25 keV. Notice, that the regions 4 and 5 were fitted simultaneously.
 
The  single {\sc vapec} model did not provide a satisfactory 
fit for the emission from region 6 ($\chi^2$/d.o.f. = 88/69).
The addition of a power law (PL) component improved the fit 
(see Table~\ref{t:best-fit}).
The addition of Gaussian component at 1.25 keV is not needed
to describe the spectra possibly due to a low count number.

The physical nature of the additional Gaussian component at 1.25 keV is not clear.
The central energy formally corresponds to Mg XI K-lines, 
but variation of the Mg abundance did not remove the excess in residuals.
Alternatively, it can be interpreted as emission of Fe-L lines 
originating from energy levels with high principal
quantum numbers \citep[e.g.][]{brickhouse2000,Yamaguchi25112011}.
This is supported by a positive correlation between 
line equivalent width $\propto K/$VEM, where $K$ is the Gaussian normalization and VEM is the volume emission measure (see Table~\ref{t:best-fit}),
and the abundance of Fe found in different regions, see Table~\ref{t:best-fit}.
While sufficient number of transitions is included in the current release of the {\sc atomdb} (v.3.0.3) used for the {\sc vapec} model, it is possible that the description of the Fe-L complex is still incomplete.


\section{Analysis of J1726}
\label{sec:j1726}

\subsection{Timing}

For the timing analysis of J1726, we used EPIC-pn data with the time resolution of 73.4 ms.
We extracted events from the 30 arcsec radius aperture in 0.3--10 keV range and 
corrected them to the solar system barycentre using {\sc barycen} tool 
and J1726 coordinates obtained with {\sc edetect\_chain} 
task (RA = $261\fdg7334$, Dec = $-38\fdg3595$).
$Z^2_n$ test \citep{ztest1983} was used to search for pulsations 
with periods in the 0.15--4000 s range.
The number of harmonics $n$ was varied from 1 to 5.
No pulsations were found.
An upper limit for the pulsed fraction (PF) of 27\% 
(at the 99\% confidence level) was estimated 
following \citet{brazier1994} and adopting $n=1$. 



\begin{figure}
\begin{minipage}[h]{1.0\linewidth}
\center{\includegraphics[width=0.67\linewidth, angle=-90]{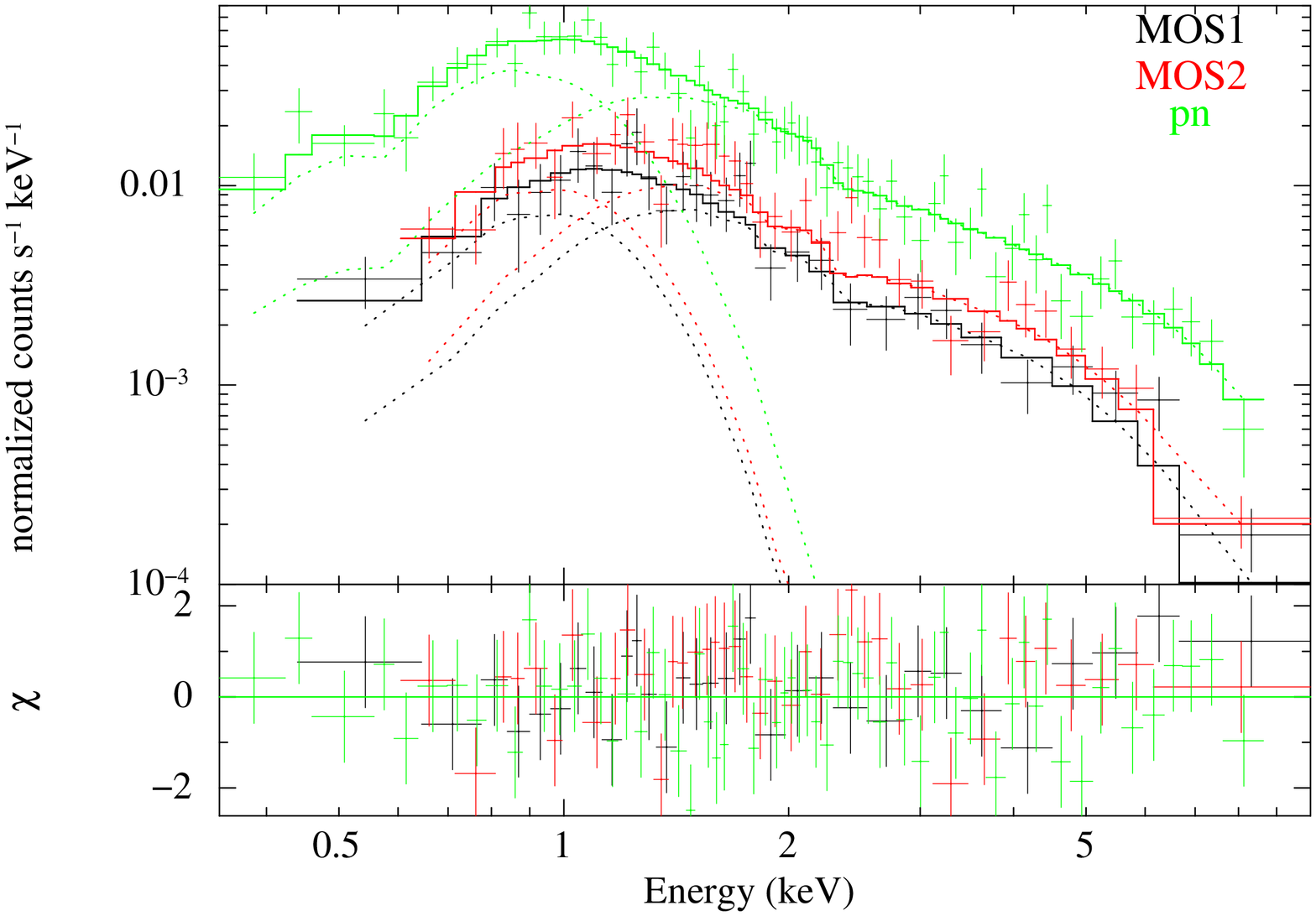}}
\end{minipage}
\begin{minipage}[h]{1.0\linewidth}
\center{\includegraphics[width=0.67\linewidth, angle=-90]{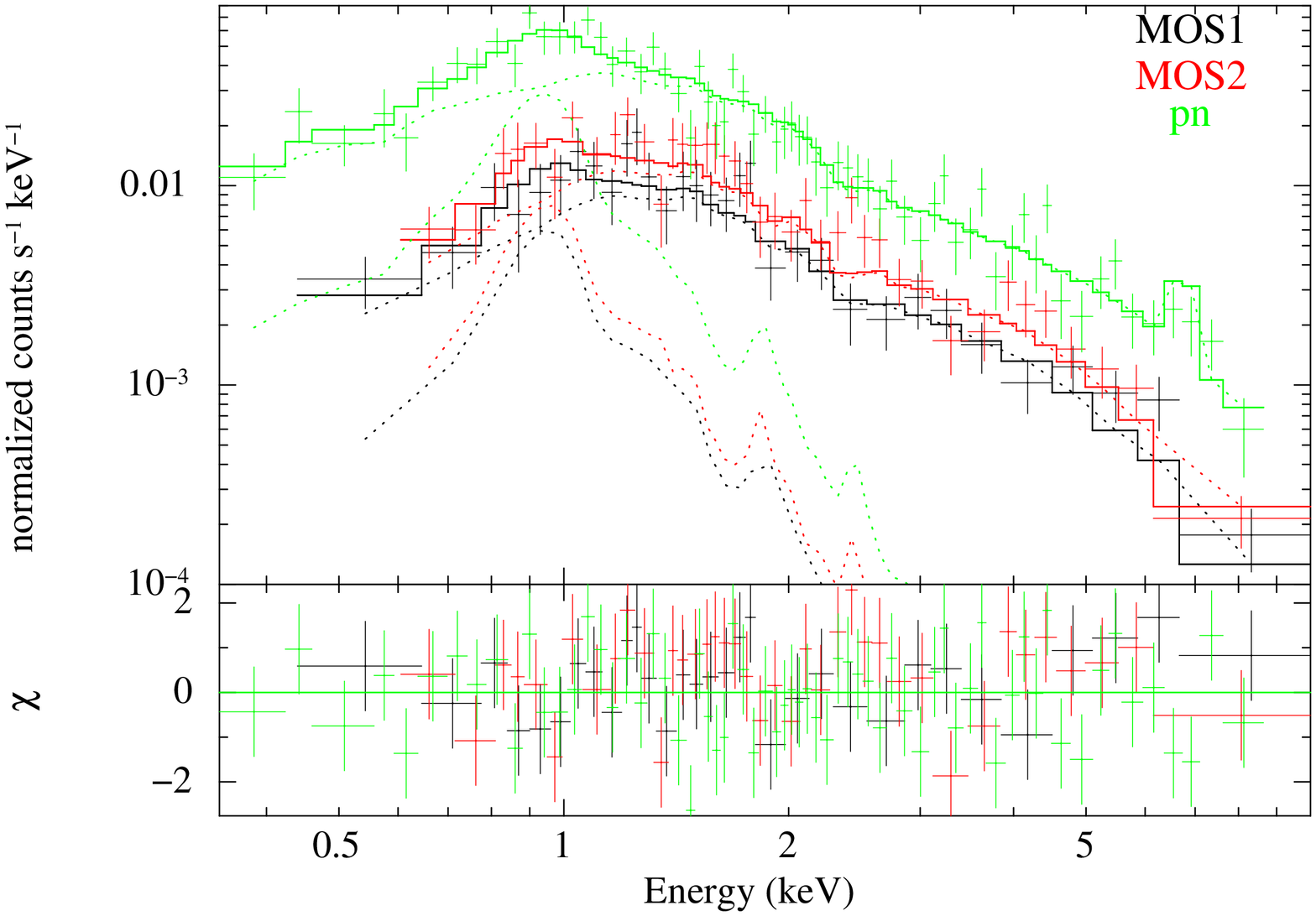}}
\end{minipage}
\caption{\textit{XMM-Newton} EPIC spectra of the J1726
best-fitted with PL+BB (top) and 2-T {\sc mekal} (bottom) models and fit residuals.
Different instruments are shown by different colors as
indicated in the top right corners. Dotted lines represent model components.}
\label{fig:j1726spec}
\end{figure}

\subsection{X-ray spectra}

We extracted the J1726 spectra from MOS1, MOS2 and pn data
from the 30 arcsec radius aperture using {\sc evselect} tool.
The 45--75 arcsec annulus region around J1726 was chosen for the background extraction. 
The resulting number of source counts was 410(MOS1)+625(MOS2)+1340(pn).
{\sc sas} tasks {\sc rmfgen} and {\sc arfgen} were utilized 
to generate the redistribution matrix and ancillary response files.
Spectra were grouped to ensure $>15$ counts per energy bin  
and then fitted simultaneously in the 0.3--10 keV range.

Assuming the NS origin of J1726, we applied the absorbed power law (PL),
power law plus blackbody (PL+BB)
and power law plus the magnetized hydrogen atmosphere \citep[PL+NSA;][]{pavlov1995} models.
The PL component describes the magnetospheric emission while BB or NSA refer to
the thermal emission from the NS surface.
We also tried the absorbed two-temperature optically thin 
thermal plasma ({\sc mekal}) model \citep*{mewe1985}
which is usually used to describe spectra of cataclysmic variables 
(CVs) \citep[see, e.g.,][]{baskill2005,reis2013}. 
The cooler component describes the emission from the 
unshocked accretion flow and white dwarf photosphere
while the hotter one describes the shocked flow emission.
The results are presented in Table~\ref{t:fit-j1726}.
According to $\chi^{2}$ values, all models describe the data well. 

The F-test was applied to investigate whether 
the thermal component is required to describe the soft part of the spectra if J1726 is a NS.
It gave a probability of chance improvement of $\approx 10^{-6}$,
showing that the thermal component is needed.
The fits for the PL+BB and 2-T {\sc mekal} models are shown in Fig.~\ref{fig:j1726spec}.
The derived photon index $\Gamma$ and the BB temperature are typical for pulsar emission.
For the distance of 3.7 kpc, the emitting area radius is consistent 
within uncertainties with a canonical NS radius of about 15 km. 
At the same time, the hydrogen atmosphere model NSA 
with NS mass $M_{\rm NS}=1.4M_\odot$ and radius $R_{\rm NS}=13$~km
can equally well describe the thermal component.   

For the 2-T {\sc mekal} model, the obtained temperatures of 0.8 and 8.6 keV are typical for 
CVs \citep[see e.g.][]{heinke2005}. 
The single temperature {\sc mekal} model was rejected by the worse fit
($\chi^2$/d.o.f.=157/135). 

\begin{landscape}

\begin{table}
\caption{J1726 best-fit spectral parameters for different models. 
All errors correspond to 90\% confidence intervals. 
Unabsorbed fluxes $f_{\rm X}$ refer to the 0.3--10 keV energy range.
For the PL+NSA model, the magnetic field $B=10^{12}$~G, the NS mass $M_{\rm NS}=1.4M_\odot$, 
the NS radius $R_{\rm NS}=13$ km, the  
distance to the object $D$ was fixed at 3.7~kpc and the temperature 
is given as seen by a distant observer. 
$N_{\rm PL}$ is the PL normalization component.}
\label{t:fit-j1726}
\begin{center}
\begin{tabular}{lccccccccc}
\hline
Parameters/ & $N_{\rm H}$,        & $\Gamma$ & $N_{\rm PL}$,                                  & $T$, & $R$, & $T_1$, & $T_2$, & $f_{\rm X}$,                   & $\chi^2$/d.o.f. \\
Model       & 10$^{21}$ cm$^{-2}$ &          & 10$^{-5}$ ph keV$^{-1}$~cm$^{-2}$~s$^{-1}$ & eV            & km   & keV   & keV   & 10$^{-13}$ erg s$^{-1}$ cm$^{-2}$ \\
\hline
\multicolumn{10}{c}{ }  \\
PL & $2.4^{+0.4}_{-0.3}$ & $1.81^{+0.12}_{-0.11}$ & $7.8^{+1.0}_{-0.9}$ & & &  & & 5.0 & 146/135 \\
\hline
\multicolumn{10}{c}{ }  \\
PL+BB & $5.1^{+2.0}_{-1.7}$ & $1.71^{+0.20}_{-0.20}$ & $7.3^{+2.3}_{-1.8}$ & $138^{+42}_{-25}$ & $5^{+10}_{-3}$ $D_{\rm 3.7kpc}$ & & & 10.2 & 119/133 \\
\hline
\multicolumn{10}{c}{ }  \\
PL+NSA & $4.3^{+0.4}_{-0.4}$ & $1.60^{+0.14}_{-0.14}$ & $6.1^{+1.1}_{-1.0}$ & $73^{+3}_{-3}$ & 13 (fixed) & & & 8.3 & 122/134 \\
\hline
\multicolumn{10}{c}{ }  \\
2-T {\sc mekal} & $2.0^{+0.4}_{-0.3}$ & & & & & $0.8^{+0.3}_{-0.1}$ & $8.6^{+4.1}_{-2.1}$ & 4.6 & 126/133 \\
\hline
\end{tabular}
\end{center}
\end{table}

\begin{figure}
\begin{minipage}[h]{0.32\linewidth}
\center{\includegraphics[scale=0.392,clip]{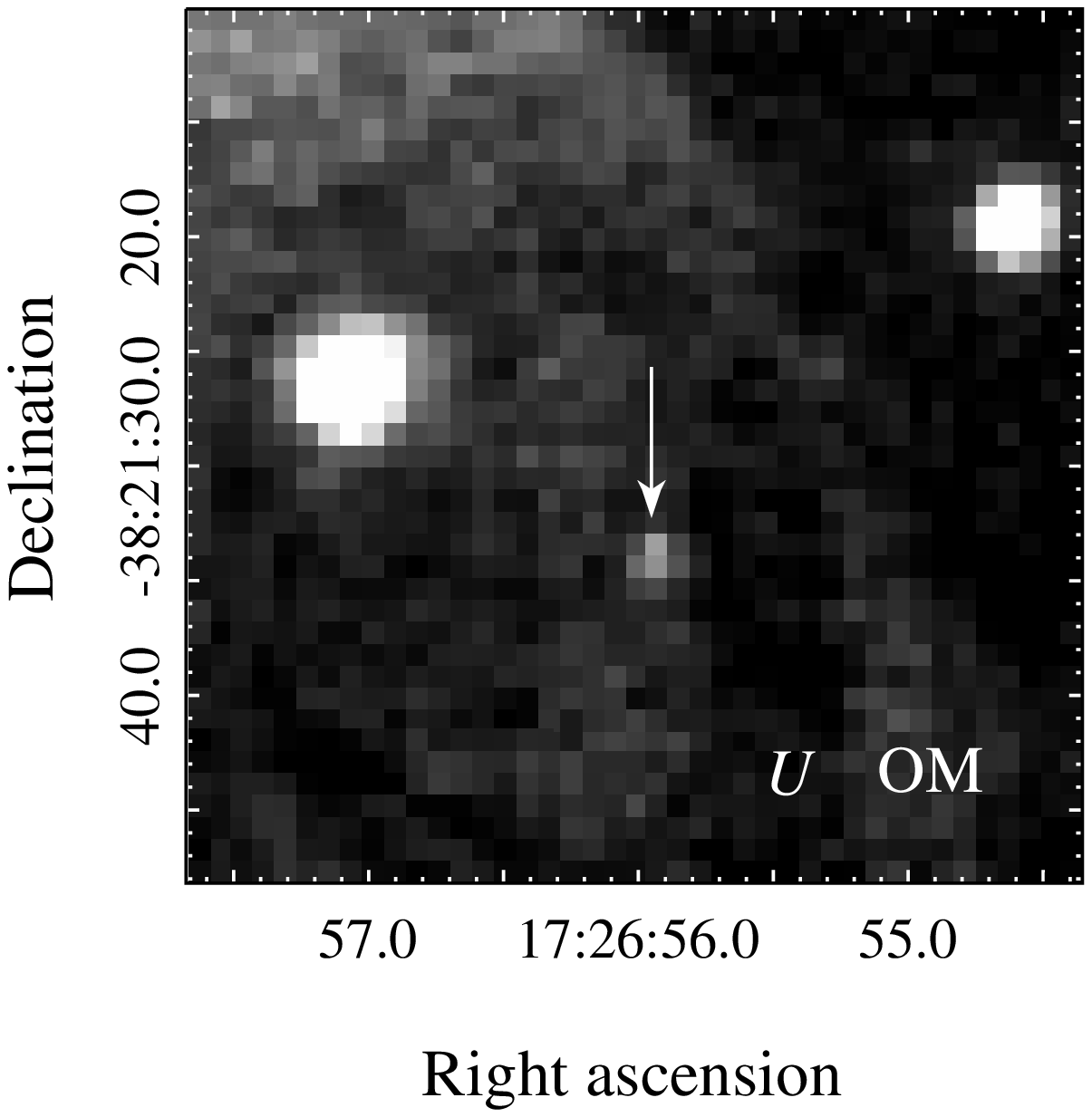}}
\end{minipage}
\begin{minipage}[h]{0.32\linewidth}
\center{\includegraphics[scale=0.392,clip]{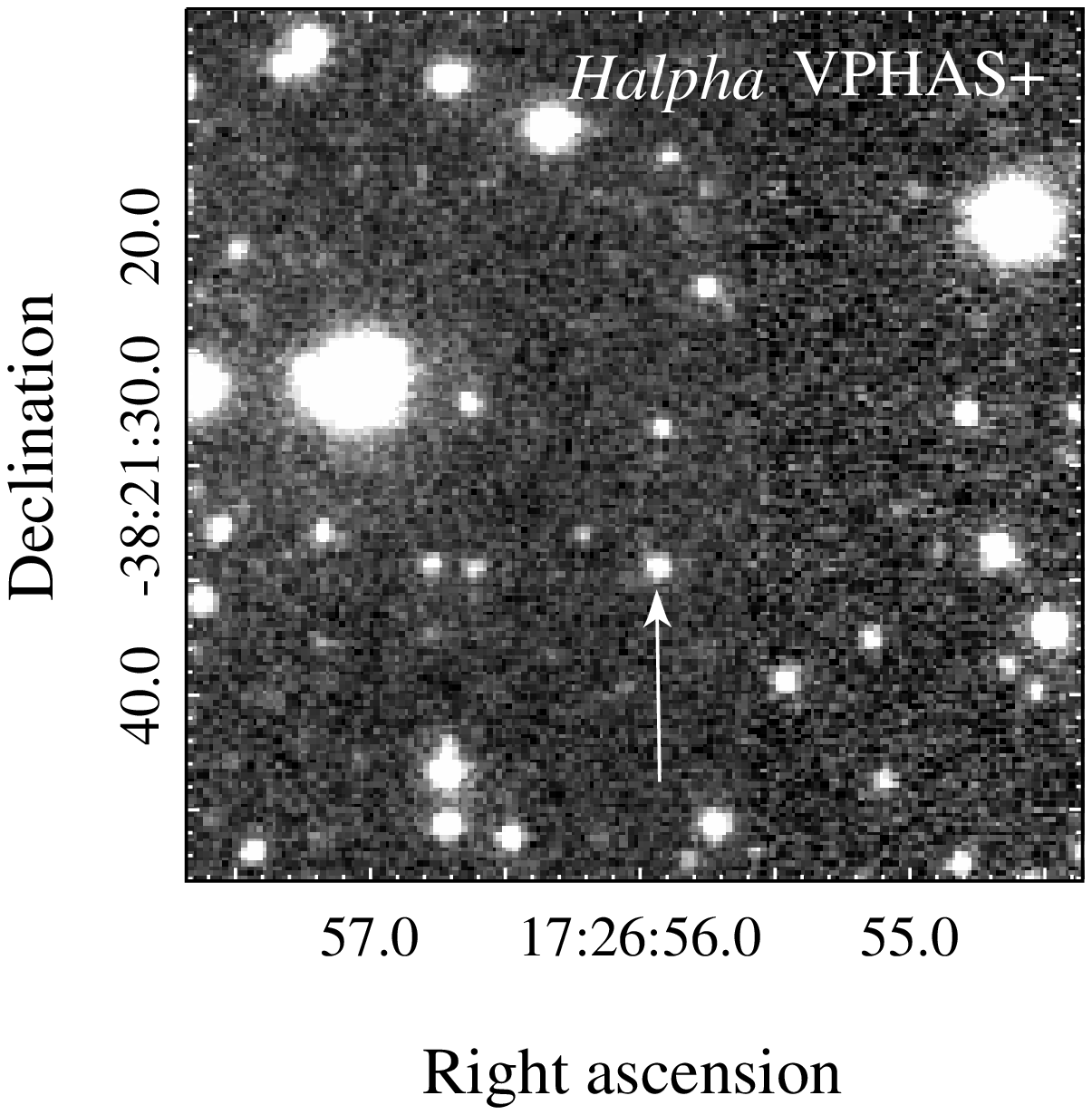}} 
\end{minipage}
\begin{minipage}[h]{0.32\linewidth}
\center{\includegraphics[scale=0.392,clip]{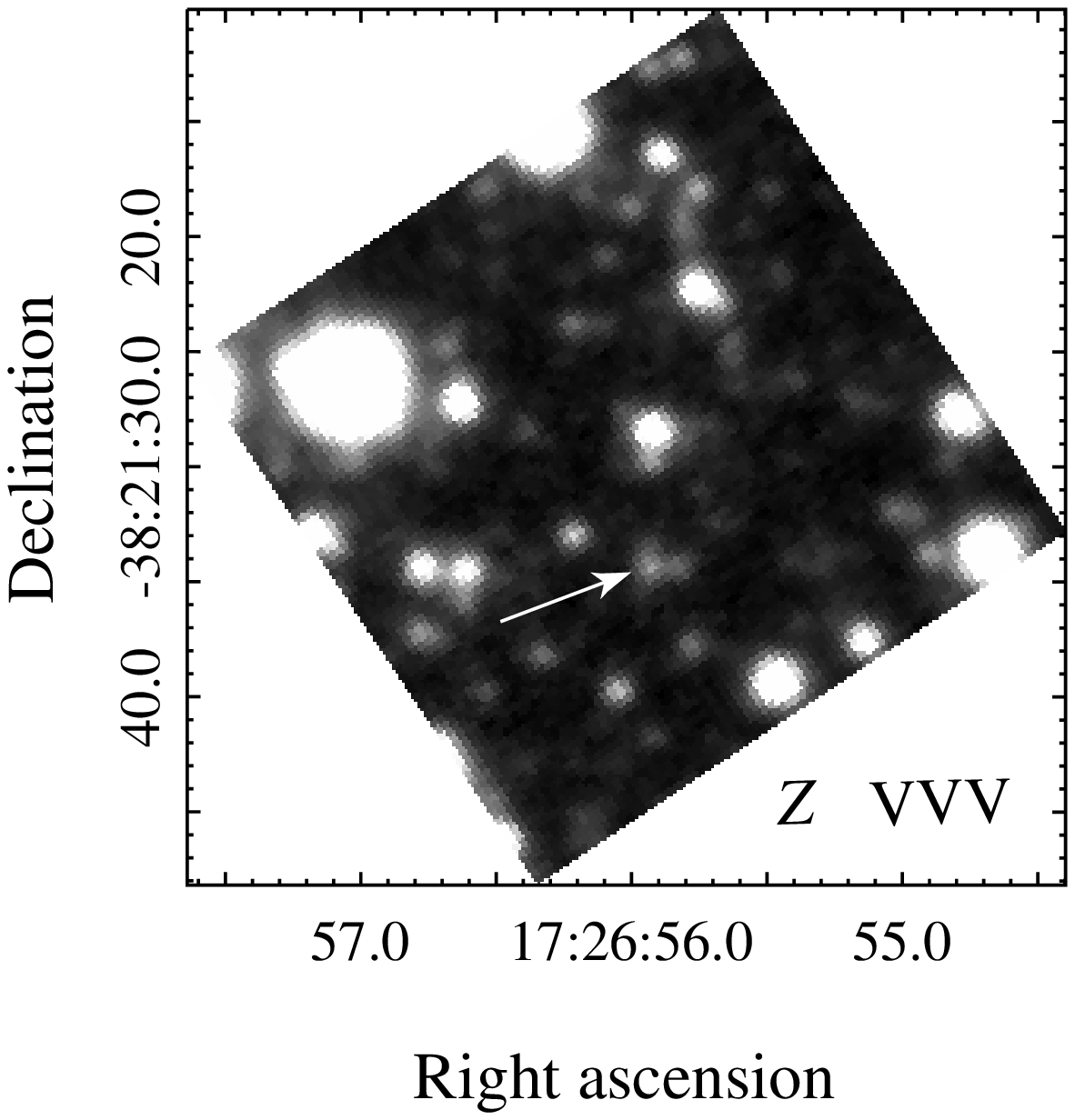}}
\end{minipage}
\caption{0.65 arcmin$\times$0.65 arcmin optical/infrared images of the J1726 field.
Filters and surveys are indicated in the images.
The J1726 possible counterpart is marked by an arrow in each image.}
\label{fig:om_halpha}
\end{figure}

\end{landscape}

\subsection{UV and optical data}

A faint optical source was detected at the J1726 position
with the \textit{XMM-Newton} Optical/UV Monitor (OM) (see Fig.~\ref{fig:om_halpha}, left panel) 
in the $U$ filter ($\lambda_{\rm eff}=344$ nm).
It can be the J1726 counterpart.
Using the {\sc omdetect} task, we obtained the 
background-subtracted count rate of 0.15 cts~s$^{-1}$ 
that corresponds to the instrumental magnitude $U = 20.35\pm0.19$ ($U_{AB} \approx 21.28\pm0.19$).
We note that this value should be considered with caution
since the source was on top of a prominent arc of straylight artefact
seen as diffuse emission in Fig.~\ref{fig:om_halpha}
that prevents to measure its count rate accurately
(\textit{XMM-Newton} support team, private communication).
The J1726 field was also observed in $V$, $UVM2$ and $UVW1$ filters
but the source was not detected in these bands.
  
At the same time, the source was found in archival images
obtained in the VISTA Variables in the Via Lactea Survey \citep[VVV; PI Dante Minniti;][]{vvv2010} 
and the VST Photometric H$\alpha$ Survey of the Southern Galactic Plane and Bulge 
\citep[VPHAS+; PI Janet Drew;][]{vphas2014}\footnote{http://www.eso.org/sci/observing/PublicSurveys.html}.
In these surveys, the source is catalogued as VPHASDR2 J172655.9$-$382134.4 and VVV J172655.88$-$382134.20.
Examples of VPHAS+ H$\alpha$ and VVV $Z$ images are shown in the 
middle and right panels of Fig.~\ref{fig:om_halpha}, respectively.
The source coordinates obtained in different surveys are 
fully consistent with the X-ray position of J1726 obtained with 
the nominal \textit{XMM-Newton} pointing accuracy of 2 arcsec 
(90\% confidence)\footnote{See http://xmm2.esac.esa.int/docs/documents/CAL-TN-0018.pdf}. 

Visual magnitudes and observed fluxes measured in different bands are presented in Table~\ref{t:flux}. $Z$, $Y$, $J$, $H$ and $K_s$ magnitudes were obtained from VVV catalogue for 1 arcsec aperture. Since the field is very crowded in the infrared band (see Fig.~\ref{fig:om_halpha}, right panel), we checked the VVV catalogue values using {\sc iraf}\footnote{{\sc iraf} is distributed by the National Optical Astronomy Observatories, which are operated by the Association of Universities for Research in Astronomy, Inc., under cooperative agreement with the National Science Foundation.}. The derived values were in agreement with the catalogue ones  within uncertainties. Nevertheless, the source emission is contaminated by nearby stars and flux values may be overestimated. VPHAS+ catalogue provides $u$, $g$, $r$, H$\alpha$ and $i$ magnitudes obtained using point spread function (PSF) fitting and aperture photometry. Both methods give consistent results. The former values are presented in Table~\ref{t:flux}. The source of interest was observed at two epochs. At MJD 56566, observations were performed in $u$, $g$ and $r$ filters and at MJD 56149 -- in $r$, H$\alpha$ and $i$ filters. In addition, two scans, primary and duplicate, are provided for each epoch. As noted in the catalogue description, the primary detection is the one for which the magnitude could be measured successfully in the largest number of filters.

\section{Discussion}
\label{sec:discus}

\subsection{The remnant}

The spectra of the most part of the SNR can be fitted with {\sc vapec+gauss} model
except for the region 6 where the additional PL component is required.
As can be seen from Table.~\ref{t:best-fit}, temperatures and absorption column densities 
obtained for different parts of the SNR  are consistent with each other within uncertainties.
Plasma in regions 1--2 has solar abundances suggesting that the emission
comes from the shocked interstellar medium 
that is typical for MM SNRs \citep{rho1998}.
However, the analysis of bright regions 3--5 revealed
the overabundance of Fe in these regions which may
indicate the presence of ejecta material.
Alternatively, the metal enrichment can be provided by the dust destruction
\citep[see e.g.][]{shelton2004}.

The measured $N_{\rm H}$ value (Table~\ref{t:best-fit}) allows to independently estimate
the distance to the remnant. The empirical $N_{\rm H}$--$A_V$ relation \citep{guver2009}
results in the optical extinction $A_V\approx3$ assuming $N_{\rm H}=6.5\times 10^{21}$~cm$^{-2}$.
Using $A_V$--distance fit from \citet*{drimmel2003}, 
we got $D\approx3$~kpc that is compatible with $D=3.7$~kpc 
obtained by \citet{case1998} from the radio surface brightness.
For the distance of 3.7 kpc, the SNR radius is about 20 pc.

The gas number density can be calculated from the {\sc vapec} model normalization given as
\begin{equation}
\label{eq:norm}
N=\frac{10^{-14}}{4\pi D^{2}_{\rm cm}}\int n_e n_{\rm H}dV\equiv  \frac{10^{-14}}{4\pi D^{2}_{\rm cm}}\times {\rm VEM},
\end{equation}
where $n_e$ and $n_{\rm H}$ are the electron and hydrogen number densities, respectively,  
$D_{\rm cm}$ is the distance in centimetres and VEM is the volume emission measure.
For solar abundances, $n_e=1.2 n_{\rm H}$ assuming almost complete ionization.
The volume of the particular X-ray emitting region was estimated as 
$V=7.3\times 10^{56}SL_{20}D_{\rm 3.7kpc}^2$ cm$^3$,
where $S$ is the area of this region in arcmin$^2$ and 
$L_{20}$ is its extension along the line of sight in the units of 20 pc.
Then
\begin{equation}
\label{eq:nh}
n_{\rm H}=13.6 N^{1/2}(SL_{20})^{-1/2}.
\end{equation}
The resulting $n_{\rm H}$ values are presented in Table~\ref{t:con} together 
with the masses of the emitting gas $M_g$ calculated assuming 
mean atomic weight for solar abundances $\mu=0.604$. 
The number density distribution seems to be rather uniform
though it depends on the volume estimation. 
The total gas mass is about 15$M_\odot$ if we assume 
identical extensions of 20 pc for all regions.

\begin{table*}
\caption{Visual magnitudes, observed and dereddened fluxes 
of the J1726 possible optical counterpart
obtained from different surveys.
Dereddening was done with the interstellar absorption 
$A_V$ = 0.9 assuming 2-T {\sc mekal} X-ray spectral model (see text for details).}
\label{t:flux}
\begin{center}
\begin{tabular}{cccccc}
\hline
Filter & $\lambda_{\rm eff}$, $\mu$m  & MJD         & Magnitude      & Flux, $\mu$Jy & Dereddened flux, $\mu$Jy\\
\hline
$U$(OM)  & 0.344                        & 56556.83866 & $20.35\pm0.19$ & $11\pm2$       & $41.7\pm7.6$        \\  
\hline
$u$    & 0.361                        & 56566.01030$^a$ & $19.81\pm0.07$ & $17.8\pm1.2$   & $65.8\pm4.4$ \\
       &                              & 56566.01254$^b$ & $19.74\pm0.07$ & $19.0\pm1.2$   & $70.3\pm4.4$ \\
\hline
$g$    & 0.468                        & 56566.02029$^a$ & $21.04\pm0.05$ & $15.7\pm0.7$   & $42.1\pm1.9$ \\
       &                              & 56566.02224$^b$ & $21.17\pm0.05$ & $13.8\pm0.7$   & $37.0\pm1.9$ \\
\hline
$r$    & 0.624                        & 56566.02705$^a$ & $20.33\pm0.06$ & $23.7\pm1.3$   & $48.6\pm2.7$ \\
       &                              & 56566.02785$^b$ & $20.51\pm0.08$ & $20.0\pm1.5$   & $41.0\pm3.1$ \\
       &                              & 56149.08459$^a$ & $20.58\pm0.06$ & $18.8\pm1.0$   & $38.6\pm2.1$ \\
       &                              & 56149.08535$^b$ & $20.66\pm0.06$ & $17.5\pm1.0$   & $35.9\pm2.1$ \\
\hline
H$\alpha$ & 0.659                    & 56149.07406$^a$ & $19.13\pm0.04$ & $55.4\pm2.1$   & $108.5\pm4.3$\\
\hline
$i$    & 0.760                        & 56149.09070$^a$ & $20.02\pm0.06$ & $25.4\pm1.4$   & $43.6\pm2.4$ \\
       &                              & 56149.09146$^b$ & $19.76\pm0.07$ & $32.1\pm2.0$   & $55.2\pm3.4$ \\   
\hline
$Z$    & 0.878                & 55725.26322 & $18.97\pm0.09$ & $58.2^{+5.2}_{-4.8}$   & $87.6^{+7.9}_{-7.3}$ \\
$Y$    & 1.021                & 55725.25799 & $19.05\pm0.13$ & $50.3^{+6.5}_{-5.8}$   & $78.3^{+9.0}_{-8.0}$ \\ 
$J$    & 1.254                & 55309.36337 & $18.55\pm0.13$ & $58.9^{+7.7}_{-6.8}$   & $74.3^{+9.8}_{-8.6}$ \\ 
$H$    & 1.646                & 55309.35370 & $17.87\pm0.17$ & $73.1^{+12.5}_{-10.7}$ & $84.9^{+14.5}_{-12.4}$ \\ 
$K_s$  & 2.149                & 55309.35857 & $17.51\pm0.20$ & $66.9^{+13.3}_{-11.1}$ & $73.8^{+14.7}_{-12.2}$ \\  
\hline
\end{tabular}
\begin{tablenotes}
\item $^a$ Primary observation.
\item $^b$ Duplicate observation.
\end{tablenotes}
\end{center}
\label{fluxes}
\end{table*}

\begin{table}
\caption{The hydrogen number densities $n_{\rm H}$ and masses of the emitting gas $M_g$ for
the different SNR regions.}
\label{t:con}
\begin{center}
\begin{tabular}{ccc}
\hline
Region & $n_{\rm H}$,  & $M_g$,  \\
       & 10$^{-2}L_{20}^{-1/2}$ cm$^{-3}$ & $L_{20}^{1/2}D_{\rm 3.7kpc}^{2}M_\odot$ \\
\hline
\multicolumn{3}{c}{}\\
1      & $4.7^{+0.4}_{-0.3}$ & $6.2^{+0.5}_{-0.4}$\\
\multicolumn{3}{c}{}\\
2      & $5.9^{+0.3}_{-0.3}$ & $6.5^{+0.3}_{-0.3}$\\
\multicolumn{3}{c}{}\\
3      & $5.4^{+0.7}_{-0.5}$ & $0.9^{+0.1}_{-0.1}$\\
\multicolumn{3}{c}{}\\
4+5   & $5.4^{+0.6}_{-0.5}$ & $1.1^{+0.1}_{-0.1}$\\
\multicolumn{3}{c}{}\\
6      & $5.4^{+1.8}_{-1.2}$ & $0.3^{+0.1}_{-0.1}$\\
\hline
\end{tabular}
\end{center}
\end{table}

\G350~has a complicated non-spherically symmetric morphology 
which is similar to that of the SNR G166.0+4.3 \citep[see e.g.][]{bocchino2009}. 
Both SNRs show three radio emitting arcs while the X-ray emission 
fills the part of the volume enclosed within two arcs. 
The G166.0+4.3 radio morphology was explained \citet{pinealt1987}
as follows. The supernova explodes in a moderately dense medium and
then the SN shock passes through a low-density cavity (hot tunnel).
\citet{gaensler1998} suggested the similar model for \G350.
The inner arc is then formed at the boundary of the cavity.
This implies that the observed X-ray emission fills the low-density region,
in qualitative accordance with the $n_{\rm H}$ values in Table~\ref{t:con}.

The spatially-resolved spectral analysis showed the uniform 
temperature distribution over the remnant. 
This is consistent with the predictions of the thermal 
conduction model \citep[e.g.][]{cox1999}.  
The conduction timescale can be estimated as
\begin{equation}
t_{\rm cond}\sim 27\frac{n_e}{1\ {\rm cm}^{-3}}\left(\frac{l_T}{10\ {\rm pc}}\right)^2\left(\frac{T}{0.8\ {\rm keV}}\right)^{-2.5}\frac{{\rm ln}\Lambda}{33}\ {\rm kyr},
\end{equation}
where $l_T=T/\nabla{T}$ is the scale length of the temperature gradient 
and ln$\Lambda=29.7+{\rm ln}n_e^{-0.5}(T/0.086\ {\rm keV})$ is the Coulomb logarithm.
For $l_T\approx 20$~pc,  $t_{\rm cond}\sim 7$~kyr is less than the estimated Sedov age of \G350~\citep{helfand1980,clark1975}.
Therefore thermal conduction may play a role in smoothing the temperature distribution. 
Direct comparison with results obtained by \citet{cox1999} is
difficult since their solutions are constructed assuming
spherical symmetry which is not the case of \G350.

There exist simulations of the X-ray emission for MM SNRs 
evolving in the non-uniform density medium. 
For example, \citet{schneiter2006} presented simulations for the SNR 3C~400.2 assuming
the SNR evolution in the medium with a jump in the density and taking into
account the effects of thermal conduction and interstellar absorption.
Their results for the SNR explosion at the denser side or at the density interface
show the X-ray emission pattern remarkably similar to that observed in \G350~(see their Fig.~4).
Thus, the \G350~morphology may be qualitatively explained assuming its evolution 
in a multi-component interstellar medium.
However, detailed numerical simulations are required.

The cloudlet evaporation model of \citet{white1991} is also
frequently used to explain MM SNR properties.
This model predicts the radial density gradient
which is not observed for \G350~(Table \ref{t:con}).
However, this cannot be considered as a solid argument 
since the model predictions for the non-spherically 
symmetric case can be different.

\subsection{The region 6}

The emission from the region 6 is harder than that from the rest of the SNR
and the additional power law component is needed to describe its spectrum.
Two weak point sources are presented inside the region 6.
The first one (RA = $261\fdg790$, Dec = $-38\fdg478$) 
shows a softer emission (it is not seen in 3--7 keV image).
A possible optical counterpart was found for this source in the USNO-B1.0 
catalogue (ID 0515$-$0517969), OM, VPHAS+ and VVV images.
The second source (RA = $261\fdg794$, Dec = $-38\fdg485$) 
has hard emission and does not have an optical counterpart.
It is possible that the PL emission results from the
incomplete  subtraction of this source by {\sc cheese} task.
We tried to use larger aperture of 30 arcsec to mask the source 
but this did not remove the PL component.
On the other hand, extracting spectra of the region 6 including point sources\footnote{
The small distance between the sources does not allow to 
separate their emission unambiguously.}
and fitting them with {\sc vapec}+PL model resulted in a similar
$\Gamma=1.9\pm0.5$ and a larger PL normalization $N_{\rm PL}=(4.3\pm 2.5)\times 10^{-5}$ ph cm$^{-2}$ s$^{-1}$ keV$^{-1}$ than in previous case (Table~\ref{t:best-fit}).
The unabsorbed 0.3--10 keV flux in the PL component changed from 1.4$\times$10$^{-13}$ to 
2.6$\times$10$^{-13}$ erg~cm$^{-2}$~s$^{-1}$. For $D=3.7$~kpc, these numbers correspond to the X-ray luminosity in the range of $(2-5)\times 10^{32}$~erg~s$^{-1}$. 
The obtained PL parameters are typical for pulsar+pulsar wind nebula (PWN)
systems with ages $\gtrsim 10$ kyr \citep{kargaltsev2008}.
This together with the spatial extent of the region~6 of $1-2$ arcmin allows us to suggest it as 
a PWN candidate. 
However, we do not see any diffuse emission that resembles a PWN in the
VLA 1.4 GHz image though it may be blended with the emission of the inner arc.

As an alternative, to check if the hard emitting component 
in the region 6 may have a thermal origin, 
we fitted its spectrum with a {\sc vapec}$+${\sc vapec} model. 
The fit is acceptable with $\chi^2$/d.o.f. = 75/67
and suggests the presence of warm and hot plasma components. 
Within uncertainties, the warm {\sc vapec} component has the same 
parameters as in the {\sc vapec}$+$PL case,
while the best fit temperature for the hot component is about 12 keV
with a 90\% lower limit of 3.6 keV. 
This is atypical for SNRs where the second hot thermal component,
if observed, has a temperature of $\la$ 3 keV \citep[e.g.][]{kawasaki2005} 
and makes the thermal interpretation of the hard component less plausible
especially for an evolved SNR such as \G350.

\subsection{J1726}

\begin{figure}
\begin{minipage}[h]{0.5\linewidth}
\center{\includegraphics[scale=0.38,clip]{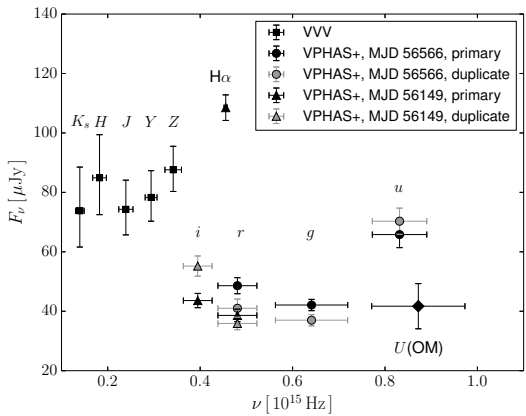}}
\end{minipage}
\caption{Dereddened flux density vs. frequency for the J1726 possible counterpart
taken from Table~\ref{t:flux}.
Data points for different observations are shown by different symbols
as indicated in the inset.}
\label{fig:opt_spec}
\end{figure}

The J1726 X-ray spectrum is well described by PL+BB(NSA) model 
with parameters typical for rotation powered pulsars. 
The corresponding column density is consistent within uncertainties
with that of the SNR supporting the NS origin of the source.
The detection of pulsations could confirm this,
but pulsations were not found (Section~\ref{sec:j1726}).
The derived upper limit for PF of 27\% is non-informative since PFs of many NSs are lower than this value.
Moreover, the actual period for the putative NS can be 
smaller than the Nyquist limit for our observations of $\approx 150$ ms. 
Higher time resolution observations are required to solve this problem. 

An optical source with a non-stellar spectral energy distribution 
was found at the X-ray position of J1726.
Assuming that the optical source is the J1726 counterpart, 
we calculated the X-ray to optical flux ratio $f_{\rm X}/f_{g}$.
The magnitude $g$ was corrected for interstellar extinction adopting
the extinction law of \citet*{cardelli1989} and $A_V=3$ (see Subsection 5.1).
The unabsorbed X-ray flux was obtained in the 0.3--10 keV energy 
band (Table~\ref{t:fit-j1726}).
We got $f_X/f_{g} \approx 1.5$. 
This is less than the values obtained for low-mass X-ray binaries ($\sim 10^2-10^3$)
or isolated NSs \citep[$\gtrsim 10^3$;][]{universeinxrays}.
While a chance spatial coincidence of a NS and an unrelated optical source can not be ruled out,
the NS interpretation seems to be unlikely. 

On the other hand, the spectral energy distribution of the putative optical counterpart 
points to the CV interpretation.
This is also supported by J1726 X-ray spectrum,
which is well described by the two-temperature optically thin plasma model
typical for CVs \citep{baskill2005,reis2013}.
The dereddened optical fluxes obtained in different bands  
are presented in the last column of Table~\ref{t:flux} and 
shown in Fig.~\ref{fig:opt_spec}.
The dereddening was performed using $A_V=0.9$ which corresponds to $N_{\rm H}$=2$\times$10$^{21}$ cm$^{-2}$ 
obtained from the 2-T {\sc mekal} fit.
The optical source shows a flux excess in a narrow-band H$\alpha$ filter (Table~\ref{t:flux})
that is a common feature for CVs \citep[see, e.g.,][]{witham2006}.
The X-ray to optical flux ratio $f_X/f_{g} \approx 7$ is also usual for CVs \citep{palombara2006}.

Other interpretations of the J1726 nature are less plausible.
For instance, J1726 cannot be an AGN since $N_{\rm H}$ obtained for the single PL fit, typical for AGNi, 
is much smaller than the total Galactic $N_{\rm H}$ in this direction of 
$9.5\times10^{21}$ cm$^{-2}$ derived from the H{\sc\,i} map by \citet{dickey1990}.


\section{Summary}
\label{sec:sum}

We analysed the \textit{XMM-Newton} observations of the Galactic MM SNR \G350.
We showed that the spectrum of the most part of the SNR 
is well described by the {\sc vapec} model with solar abundances
suggesting that the X-ray emission originates from the shocked interstellar material.
However, the analysis of some bright regions revealed an overabundance of Fe
possibly indicating the presence of ejecta.
No temperature variations over the observed part of the remnant were found
indicating that thermal conduction effects may be important.
The complicated multiwavelength morphology of \G350~can be explained assuming its
evolution in the medium with jumps in the density.
Deeper X-ray observations are required to investigate the properties of the entire SNR.
The X-ray spectra of the brightest source J1726 in the SNR field 
can be fitted either by PL+BB(NSA) or 2-T {\sc mekal} models.
In the former case, the best-fit parameters are
typical for a rotation powered pulsar and
the absorption column density is in agreement with the SNR one
although no pulsed emission was detected with PF upper limit of 27\%.
Alternatively, J1726 is a foreground source, likely a CV.
The latter possibility is supported by a 
faint optical source with a non-stellar spectrum found 
at the J1726 X-ray position. 
It is possible that one of other point sources detected in the SNR field 
can be an associated compact object, 
however current X-ray data do not allow to reliably define their nature.
For instance, the source with hard emission in region 6 
is promising in this respect. 
Deeper X-ray observations with better spatial and timing resolution 
are needed to investigate it more accurately.

\section*{Acknowledgments}
We thank the anonymous referee for useful comments
which helped us to improve the paper.
The scientific results reported in this article are based on 
observations obtained with XMM-Newton, an ESA science mission 
with instruments and contributions directly 
funded by ESA Member States and the USA (NASA).
We thank Sergey Zharikov for helpful discussions.
YS is grateful to Russian Foundation for 
Basic Research for support (project \# 14-02-00868).
The work of AD was supported by Russian Foundation for Basic Research 
 under research project \# 16-32-00504 mol\_a.

\bibliographystyle{mnras}
\bibliography{refG350}

\begin{thebibliography}{}
\makeatletter
\relax
\def\mn@urlcharsother{\let\do\@makeother \do\$\do\&\do\#\do\^\do\_\do\%\do\~}
\def\mn@doi{\begingroup\mn@urlcharsother \@ifnextchar [ {\mn@doi@}
  {\mn@doi@[]}}
\def\mn@doi@[#1]#2{\def\@tempa{#1}\ifx\@tempa\@empty \href
  {http://dx.doi.org/#2} {doi:#2}\else \href {http://dx.doi.org/#2} {#1}\fi
  \endgroup}
\def\mn@eprint#1#2{\mn@eprint@#1:#2::\@nil}
\def\mn@eprint@arXiv#1{\href {http://arxiv.org/abs/#1} {{\tt arXiv:#1}}}
\def\mn@eprint@dblp#1{\href {http://dblp.uni-trier.de/rec/bibtex/#1.xml}
  {dblp:#1}}
\def\mn@eprint@#1:#2:#3:#4\@nil{\def\@tempa {#1}\def\@tempb {#2}\def\@tempc
  {#3}\ifx \@tempc \@empty \let \@tempc \@tempb \let \@tempb \@tempa \fi \ifx
  \@tempb \@empty \def\@tempb {arXiv}\fi \@ifundefined
  {mn@eprint@\@tempb}{\@tempb:\@tempc}{\expandafter \expandafter \csname
  mn@eprint@\@tempb\endcsname \expandafter{\@tempc}}}

\bibitem[\protect\citeauthoryear{{Anders} \& {Grevesse}}{{Anders} \&
  {Grevesse}}{1989}]{angr1989}
{Anders} E.,  {Grevesse} N.,  1989, \mn@doi [\gca]
  {10.1016/0016-7037(89)90286-X}, \href
  {http://adsabs.harvard.edu/abs/1989GeCoA..53..197A} {53, 197}

\bibitem[\protect\citeauthoryear{{Baskill}, {Wheatley}  \& {Osborne}}{{Baskill}
  et~al.}{2005}]{baskill2005}
{Baskill} D.~S.,  {Wheatley} P.~J.,   {Osborne} J.~P.,  2005, \mn@doi [\mnras]
  {10.1111/j.1365-2966.2005.08677.x}, \href
  {http://adsabs.harvard.edu/abs/2005MNRAS.357..626B} {357, 626}

\bibitem[\protect\citeauthoryear{{Bocchino}, {Miceli}  \& {Troja}}{{Bocchino}
  et~al.}{2009}]{bocchino2009}
{Bocchino} F.,  {Miceli} M.,   {Troja} E.,  2009, \mn@doi [\aap]
  {10.1051/0004-6361/200810742}, \href
  {http://adsabs.harvard.edu/abs/2009A%26A...498..139B} {498, 139}

\bibitem[\protect\citeauthoryear{{Brazier}}{{Brazier}}{1994}]{brazier1994}
{Brazier} K.~T.~S.,  1994, \mnras, \href
  {http://adsabs.harvard.edu/abs/1994MNRAS.268..709B} {268, 709}

\bibitem[\protect\citeauthoryear{{Brickhouse}, {Dupree}, {Edgar}, {Liedahl},
  {Drake}, {White}  \& {Singh}}{{Brickhouse} et~al.}{2000}]{brickhouse2000}
{Brickhouse} N.~S.,  {Dupree} A.~K.,  {Edgar} R.~J.,  {Liedahl} D.~A.,  {Drake}
  S.~A.,  {White} N.~E.,   {Singh} K.~P.,  2000, \mn@doi [\apj]
  {10.1086/308350}, \href {http://adsabs.harvard.edu/abs/2000ApJ...530..387B}
  {530, 387}

\bibitem[\protect\citeauthoryear{{Buccheri} et~al.}{{Buccheri}
  et~al.}{1983}]{ztest1983}
{Buccheri} R.,  et~al., 1983, \aap, \href
  {http://adsabs.harvard.edu/abs/1983A%26A...128..245B} {128, 245}

\bibitem[\protect\citeauthoryear{{Cardelli}, {Clayton}  \& {Mathis}}{{Cardelli}
  et~al.}{1989}]{cardelli1989}
{Cardelli} J.~A.,  {Clayton} G.~C.,   {Mathis} J.~S.,  1989, \mn@doi [\apj]
  {10.1086/167900}, \href {http://adsabs.harvard.edu/abs/1989ApJ...345..245C}
  {345, 245}

\bibitem[\protect\citeauthoryear{{Case} \& {Bhattacharya}}{{Case} \&
  {Bhattacharya}}{1998}]{case1998}
{Case} G.~L.,  {Bhattacharya} D.,  1998, \mn@doi [\apj] {10.1086/306089}, \href
  {http://adsabs.harvard.edu/abs/1998ApJ...504..761C} {504, 761}

\bibitem[\protect\citeauthoryear{{Caswell}, {Clark}, {Crawford}  \&
  {Green}}{{Caswell} et~al.}{1975}]{caswell1975}
{Caswell} J.~L.,  {Clark} D.~H.,  {Crawford} D.~F.,   {Green} A.~J.,  1975,
  Australian Journal of Physics Astrophysical Supplement, \href
  {http://adsabs.harvard.edu/abs/1975AuJPA..37....1C} {37, 1}

\bibitem[\protect\citeauthoryear{{Clark} \& {Stephenson}}{{Clark} \&
  {Stephenson}}{1975}]{clark1975}
{Clark} D.~H.,  {Stephenson} F.~R.,  1975, The Observatory, \href
  {http://adsabs.harvard.edu/abs/1975Obs....95..190C} {95, 190}

\bibitem[\protect\citeauthoryear{{Cox}, {Shelton}, {Maciejewski}, {Smith},
  {Plewa}, {Pawl}  \& {R{\'o}{\.z}yczka}}{{Cox} et~al.}{1999}]{cox1999}
{Cox} D.~P.,  {Shelton} R.~L.,  {Maciejewski} W.,  {Smith} R.~K.,  {Plewa} T.,
  {Pawl} A.,   {R{\'o}{\.z}yczka} M.,  1999, \mn@doi [\apj] {10.1086/307781},
  \href {http://adsabs.harvard.edu/abs/1999ApJ...524..179C} {524, 179}

\bibitem[\protect\citeauthoryear{{Dickey} \& {Lockman}}{{Dickey} \&
  {Lockman}}{1990}]{dickey1990}
{Dickey} J.~M.,  {Lockman} F.~J.,  1990, \mn@doi [\araa]
  {10.1146/annurev.aa.28.090190.001243}, \href
  {http://adsabs.harvard.edu/abs/1990ARA%26A..28..215D} {28, 215}

\bibitem[\protect\citeauthoryear{{Drew} et~al.,}{{Drew}
  et~al.}{2014}]{vphas2014}
{Drew} J.~E.,  et~al., 2014, \mn@doi [\mnras] {10.1093/mnras/stu394}, \href
  {http://adsabs.harvard.edu/abs/2014MNRAS.440.2036D} {440, 2036}

\bibitem[\protect\citeauthoryear{{Drimmel}, {Cabrera-Lavers}  \&
  {L{\'o}pez-Corredoira}}{{Drimmel} et~al.}{2003}]{drimmel2003}
{Drimmel} R.,  {Cabrera-Lavers} A.,   {L{\'o}pez-Corredoira} M.,  2003, \mn@doi
  [\aap] {10.1051/0004-6361:20031070}, \href
  {http://adsabs.harvard.edu/abs/2003A%26A...409..205D} {409, 205}

\bibitem[\protect\citeauthoryear{{Gaensler}}{{Gaensler}}{1998}]{gaensler1998}
{Gaensler} B.~M.,  1998, \mn@doi [\apj] {10.1086/305146}, \href
  {http://adsabs.harvard.edu/abs/1998ApJ...493..781G} {493, 781}

\bibitem[\protect\citeauthoryear{{G{\"u}ver} \& {{\"O}zel}}{{G{\"u}ver} \&
  {{\"O}zel}}{2009}]{guver2009}
{G{\"u}ver} T.,  {{\"O}zel} F.,  2009, \mn@doi [\mnras]
  {10.1111/j.1365-2966.2009.15598.x}, \href
  {http://adsabs.harvard.edu/abs/2009MNRAS.400.2050G} {400, 2050}

\bibitem[\protect\citeauthoryear{{Heinke}, {Grindlay}, {Edmonds}, {Cohn},
  {Lugger}, {Camilo}, {Bogdanov}  \& {Freire}}{{Heinke}
  et~al.}{2005}]{heinke2005}
{Heinke} C.~O.,  {Grindlay} J.~E.,  {Edmonds} P.~D.,  {Cohn} H.~N.,  {Lugger}
  P.~M.,  {Camilo} F.,  {Bogdanov} S.,   {Freire} P.~C.,  2005, \apj, \href
  {http://adsabs.harvard.edu/abs/2005AIPC..797...40H} {625, 796}

\bibitem[\protect\citeauthoryear{{Helfand}, {Chanan}  \& {Novick}}{{Helfand}
  et~al.}{1980}]{helfand1980}
{Helfand} D.~J.,  {Chanan} G.~A.,   {Novick} R.,  1980, \mn@doi [\nat]
  {10.1038/283337a0}, \href {http://adsabs.harvard.edu/abs/1980Natur.283..337H}
  {283, 337}

\bibitem[\protect\citeauthoryear{{Kargaltsev} \& {Pavlov}}{{Kargaltsev} \&
  {Pavlov}}{2008}]{kargaltsev2008}
{Kargaltsev} O.,  {Pavlov} G.~G.,  2008, in {Bassa} C.,  {Wang} Z.,  {Cumming}
  A.,   {Kaspi} V.~M.,  eds,  American Institute of Physics Conference Series
  Vol. 983, 40 Years of Pulsars: Millisecond Pulsars, Magnetars and More. pp
  171--185 (\mn@eprint {arXiv} {0801.2602}), \mn@doi{10.1063/1.2900138}

\bibitem[\protect\citeauthoryear{{Kaspi}, {Manchester}, {Johnston}, {Lyne}  \&
  {D'Amico}}{{Kaspi} et~al.}{1996}]{kaspi1996}
{Kaspi} V.~M.,  {Manchester} R.~N.,  {Johnston} S.,  {Lyne} A.~G.,   {D'Amico}
  N.,  1996, \mn@doi [\aj] {10.1086/117938}, \href
  {http://adsabs.harvard.edu/abs/1996AJ....111.2028K} {111, 2028}

\bibitem[\protect\citeauthoryear{{Kawasaki}, {Ozaki}, {Nagase}, {Inoue}  \&
  {Petre}}{{Kawasaki} et~al.}{2005}]{kawasaki2005}
{Kawasaki} M.,  {Ozaki} M.,  {Nagase} F.,  {Inoue} H.,   {Petre} R.,  2005,
  \mn@doi [\apj] {10.1086/432591}, \href
  {http://adsabs.harvard.edu/abs/2005ApJ...631..935K} {631, 935}

\bibitem[\protect\citeauthoryear{{La Palombara}, {Mignani}, {Hatziminaoglou},
  {Schirmer}, {Bignami}  \& {Caraveo}}{{La Palombara}
  et~al.}{2006}]{palombara2006}
{La Palombara} N.,  {Mignani} R.~P.,  {Hatziminaoglou} E.,  {Schirmer} M.,
  {Bignami} G.~F.,   {Caraveo} P.,  2006, \mn@doi [\aap]
  {10.1051/0004-6361:20065247}, \href
  {http://adsabs.harvard.edu/abs/2006A%26A...458..245L} {458, 245}

\bibitem[\protect\citeauthoryear{{Lazendic} \& {Slane}}{{Lazendic} \&
  {Slane}}{2006}]{lazendic2006}
{Lazendic} J.~S.,  {Slane} P.~O.,  2006, \mn@doi [\apj] {10.1086/505380}, \href
  {http://adsabs.harvard.edu/abs/2006ApJ...647..350L} {647, 350}

\bibitem[\protect\citeauthoryear{{Mewe}, {Gronenschild}  \& {van den
  Oord}}{{Mewe} et~al.}{1985}]{mewe1985}
{Mewe} R.,  {Gronenschild} E.~H.~B.~M.,   {van den Oord} G.~H.~J.,  1985,
  \aaps, \href {http://adsabs.harvard.edu/abs/1985A%26AS...62..197M} {62, 197}

\bibitem[\protect\citeauthoryear{{Minniti} et~al.,}{{Minniti}
  et~al.}{2010}]{vvv2010}
{Minniti} D.,  et~al., 2010, \mn@doi [New Astronomy]
  {10.1016/j.newast.2009.12.002}, \href
  {http://adsabs.harvard.edu/abs/2010NewA...15..433M} {15, 433}

\bibitem[\protect\citeauthoryear{{Pavlov}, {Shibanov}, {Zavlin}  \&
  {Meyer}}{{Pavlov} et~al.}{1995}]{pavlov1995}
{Pavlov} G.~G.,  {Shibanov} Y.~A.,  {Zavlin} V.~E.,   {Meyer} R.~D.,  1995, in
  {Alpar} M.~A.,  {Kiziloglu} U.,   {van Paradijs} J.,  eds, The Lives of the
  Neutron Stars. Kluwer, Dordrecht, p.~71

\bibitem[\protect\citeauthoryear{{Pineault}, {Landecker}  \&
  {Routledge}}{{Pineault} et~al.}{1987}]{pinealt1987}
{Pineault} S.,  {Landecker} T.~L.,   {Routledge} D.,  1987, \mn@doi [\apj]
  {10.1086/165161}, \href {http://adsabs.harvard.edu/abs/1987ApJ...315..580P}
  {315, 580}

\bibitem[\protect\citeauthoryear{{Reis}, {Wheatley}, {G{\"a}nsicke}  \&
  {Osborne}}{{Reis} et~al.}{2013}]{reis2013}
{Reis} R.~C.,  {Wheatley} P.~J.,  {G{\"a}nsicke} B.~T.,   {Osborne} J.~P.,
  2013, \mn@doi [\mnras] {10.1093/mnras/stt025}, \href
  {http://adsabs.harvard.edu/abs/2013MNRAS.430.1994R} {430, 1994}

\bibitem[\protect\citeauthoryear{{Rho} \& {Petre}}{{Rho} \&
  {Petre}}{1998}]{rho1998}
{Rho} J.,  {Petre} R.,  1998, \mn@doi [\apjl] {10.1086/311538}, \href
  {http://adsabs.harvard.edu/abs/1998ApJ...503L.167R} {503, L167}

\bibitem[\protect\citeauthoryear{{Schneiter}, {de La Fuente}  \&
  {Vel{\'a}zquez}}{{Schneiter} et~al.}{2006}]{schneiter2006}
{Schneiter} E.~M.,  {de La Fuente} E.,   {Vel{\'a}zquez} P.~F.,  2006, \mn@doi
  [\mnras] {10.1111/j.1365-2966.2006.10652.x}, \href
  {http://adsabs.harvard.edu/abs/2006MNRAS.371..369S} {371, 369}

\bibitem[\protect\citeauthoryear{{Shelton}, {Kuntz}  \& {Petre}}{{Shelton}
  et~al.}{2004}]{shelton2004}
{Shelton} R.~L.,  {Kuntz} K.~D.,   {Petre} R.,  2004, \mn@doi [\apj]
  {10.1086/422352}, \href {http://adsabs.harvard.edu/abs/2004ApJ...611..906S}
  {611, 906}

\bibitem[\protect\citeauthoryear{{Snowden} \& {Kuntz}}{{Snowden} \&
  {Kuntz}}{2014}]{cookbook}
{Snowden} S.~L.,  {Kuntz} K.~D.,  2014, {Cookbook for analysis procedures for
  XMM-Newton EPIC observations of extendedobjects and the diffuse background}

\bibitem[\protect\citeauthoryear{{Stupar} \& {Parker}}{{Stupar} \&
  {Parker}}{2011}]{stupar2011}
{Stupar} M.,  {Parker} Q.~A.,  2011, \mn@doi [\mnras]
  {10.1111/j.1365-2966.2011.18547.x}, \href
  {http://adsabs.harvard.edu/abs/2011MNRAS.414.2282S} {414, 2282}

\bibitem[\protect\citeauthoryear{{Tr{\"u}mper} \& {Hasinger}}{{Tr{\"u}mper} \&
  {Hasinger}}{2008}]{universeinxrays}
{Tr{\"u}mper} J.~E.,  {Hasinger} G.,  2008, {The Universe in X-Rays},
  \mn@doi{10.1007/978-3-540-34412-4.
}

\bibitem[\protect\citeauthoryear{{White} \& {Long}}{{White} \&
  {Long}}{1991}]{white1991}
{White} R.~L.,  {Long} K.~S.,  1991, \mn@doi [\apj] {10.1086/170073}, \href
  {http://adsabs.harvard.edu/abs/1991ApJ...373..543W} {373, 543}

\bibitem[\protect\citeauthoryear{{Witham} et~al.,}{{Witham}
  et~al.}{2006}]{witham2006}
{Witham} A.~R.,  et~al., 2006, \mn@doi [\mnras]
  {10.1111/j.1365-2966.2006.10395.x}, \href
  {http://adsabs.harvard.edu/abs/2006MNRAS.369..581W} {369, 581}

\bibitem[\protect\citeauthoryear{Yamaguchi, Koyama  \& Uchida}{Yamaguchi
  et~al.}{2011}]{Yamaguchi25112011}
Yamaguchi H.,  Koyama K.,   Uchida H.,  2011, \mn@doi [Publications of the
  Astronomical Society of Japan] {10.1093/pasj/63.sp3.S837}, 63, S837

\makeatother
\end{thebibliography}

\clearpage

\end{document}